\documentclass[10pt,a4paper,onecolumn]{IEEEtran}
\usepackage[ruled,vlined]{algorithm2e}
\usepackage{amsfonts}
\usepackage{amssymb}

\usepackage{cancel}
\usepackage{ClearSans}

\usepackage{dsfont}

\usepackage{epsfig}

\usepackage{glossaries}
\usepackage{mathrsfs}
\usepackage{mathtools}
\usepackage{mfirstuc}
\usepackage{pgfplots}

\usepackage{scalerel}
\usepackage[binary-units]{siunitx}
\usepackage{subcaption}

\usepackage{tikz}

\usetikzlibrary{arrows,decorations.markings, shapes.arrows}
\usetikzlibrary{calc,intersections, positioning,graphs,fit}
\usetikzlibrary{shapes.geometric}
\usetikzlibrary{graphs}
\usetikzlibrary{overlay-beamer-styles}
\usetikzlibrary{decorations.text}
\usetikzlibrary{chains,shadings,trees}

\pgfdeclarelayer{background}
\pgfdeclarelayer{foreground}
\pgfsetlayers{background,main,foreground}

\allowdisplaybreaks

\usepackage{enumitem}
\usepackage[draft]{hyperref}

\newcommand{\field}[1]{\mathbb{#1}}

\newcommand{\indicator}{\mathds{1}}

\newcommand{\Exp}[2][]{\ensuremath{\mathbb{E}_{#1}\left[#2 \right]}} 
\newcommand{\var}[1]{\ensuremath{{\rm var} \left[#1\right]}} 
\newcommand{\cov}[1]{\ensuremath{{\rm cov} \left[#1\right]}} 

\newcommand{\eqann}[2]{\overset{\mathclap{(\text{#2})}}{#1}} 
\newcommand{\eqannref}[1]{$(\text{#1})$}
\newcommand{\bv}[1]{\mathbf{#1}} 
\newcommand{\rv}[1]{\mathsf{#1}} 
\newcommand{\Prob}[1]{\ensuremath{\mathsf{P} \left( #1 \right)}} 

\newcommand{\minus}{\scalebox{0.75}[1.0]{\( - \)}}

\newcommand{\plus}{\scalebox{0.75}[1.0]{\( + \)}}

\newcommand{\set}[1]{\mathcal{#1}} 
\newcommand{\setrn}[2]{ \ensuremath{\mathcal{I}^{(\! #2 \!)}_{#1} }}
\newcommand{\typsetn}[2]{\mathcal{T}_{#1}^{(n)} (#2)} 

  \DeclareMathOperator*{\argmin}{arg\,min}

\newtheorem{theorem}{Theorem}
\newtheorem{corollary}{Corollary}
\newtheorem{remark}{Remark}

\newtheorem{lemma}{Lemma}

\newenvironment{proof}[1][Proof]{\noindent\textbf{#1.} }{\ \rule{0.5em}{0.5em}}

\newtheorem{proposition}{Proposition}

\newacronym{awgn}{AWGN}{Additive White Gaussian Noise}

\newacronym{bac}{BAC}{Binary Asymmetric Channel}

\newacronym{bdsib}{BDSIB}{Binary Double-Sided Information-Bottleneck}
\newacronym{bec}{BEC}{Binary Erasure Channel}

\newacronym{bms}{BMS}{Binary Memoryless Symmetric}

\newacronym{bs}{BS}{base station}

\newacronym{bsc}{BSC}{Binary Symmetric Channel}
\newacronym{bscs}{BSCs}{Binary Symmetric Channels}

\newacronym{ceb}{CEB}{Conditional Entropy Bound}

\newacronym{cr}{CR}{Common Reconstruction}

\newacronym{cran}{C-RAN}{cloud radio access network}

\newacronym{cp}{CP}{central processor}

\newacronym{dmc}{DMC}{Discrete Memoryless Channel}

\newacronym{dmmac}{DM-MAC}{Discrete Memoryless Multiple Access Channel}

\newacronym{dms}{DMS}{Discrete Memoryless Source}

\newacronym{dpi}{DPI}{Data Proccesing Inequality}

\newacronym{dsbs}{DSBS}{Doubly Symmetric Binary Source}

\newacronym{dsib}{DSIB}{Double-Sided Information-Bottleneck}

\newacronym{epi}{EPI}{Entropy Power Inequality}

\newacronym{gdsib}{GDSIB}{Gaussian Double-Sided Information Bottleneck}

\newacronym{gp}{GP}{Gelf'and-Pinsker}

\newacronym{ib}{IB}{Information Bottleneck}
\newacronym{iid}{i.i.d.}{independent and identically distributed}

\newacronym{infcomb}{IC}{Information Combining}

\newacronym{lhs}{LHS}{Left Hand Side}

\newacronym{sgdsib}{SGDSIB}{Scalar Gaussian Double-Sided Information-Bottleneck}

\newacronym{mac}{MAC}{multiple access channel}

\newacronym{mgl}{MGL}{Mrs. Gerber's Lemma}

\newacronym{mi}{MI}{mutual information}

\newacronym{mu}{MU}{Mobile User}

\newacronym{oblib}{OBLIB}{Oblivious Information Bottleneck}

\newacronym{pf}{PF}{Privacy Funnel}

\newacronym{pmf}{\textsf{pmf}}{probability mass function}

\newacronym{ssib}{SSIB}{Single-Sided Information Bottleneck}
\newacronym{sgssib}{SGSSIB}{Scalar Gaussian Single-Sided Information Bottleneck function}
\newacronym{sawgnssib}{SAWGNSSIB}{Scalar AWGN Single-Sided Information Bottleneck function}

\newacronym{snr}{SNR}{Signal to Noise Ratio}

\newacronym{tibo}{TIBO}{Ternary-Input Binary-Output }

\newacronym{tito}{TITO}{Ternary-Input Ternary-Output }

\newacronym{qiqo}{QIQO}{Quaternary-Input Quaternary-Output }

\newacronym{rhs}{RHS}{right hand side}

\newacronym{rv}{RV}{Random Variable}

\newacronym{rssib}{RSSIB}{Reversed Single-Sided Information Bottleneck function}
\newacronym{rsawgnssib}{RSAWGNSSIB}{Reversed Scalar AWGN Single-Sided Information Bottleneck function}

\newacronym{wlog}{WLOG}{long}

\newacronym{wtc}{WTC}{Wiretap Channel}

\title{Bounds on the Capacity of the Multiple Access Diamond Channel with Cooperating Base-Stations}

\author{
    \IEEEauthorblockN{Michael Dikshtein \IEEEauthorrefmark{1}, Shirin Saeedi Bidokhti \IEEEauthorrefmark{2} and Shlomo Shamai (Shitz) \IEEEauthorrefmark{1}}
    
	\IEEEauthorblockA{\IEEEauthorrefmark{1} Electrical and
		Computer Engineering,
		Technion,
		Haifa, Israel,\\
		Email: \{michaeldic@campus,sshlomo@ee\}.technion.ac.il}
		
	\IEEEauthorblockA{\IEEEauthorrefmark{2} Electrical and System Engineering,
		Universty of Pennsylvania, USA,\\
		Email: saeedi@seas.upenn.edu}
}
	
\begin{document}
	\maketitle	
	\begin{abstract}
	A diamond network is considered in which the \acrlong{cp} is connected, via backhaul noiseless links, to multiple conferencing base stations, which communicate with a single user over a \acrlong{mac}. We propose coding techniques along with lower and upper bounds on the capacity. Our achievability scheme uses a common cloud coding strategy based on the technique proposed by Wand, Wigger, and Zaidi (2018) and extends it beyond two relays.  Our upper bounds generalize the method proposed by Bidokhti and Kramer for the two relay diamond network without cooperation (2016) and lead to new bounds for the multiple relay setting. Specializing our upper bounds for the two relay scenario (with cooperation), we provide new bounds and improve state-of-the-art.
\end{abstract}

	\section{Introduction}

Cloud radio access networks (C-RANs) play a central role in enabling modern, reliable, ultra-bandwidth, scalable, and fast communication systems.
As the demand for  steadily growing data transmission increases, capacity stands as the main challenge for every emerging generation of mobile networks \cite{Bockelmann2016}. Ultra-dense cell deployment with cooperative operations will become an enabling technology for this vision \cite{Rost2014}. 
In traditional architectures, radio and baseband processing functionality is solely implemented inside a \acrfull{bs}. Conversely, in \acrshort{cran}, the \acrshort{bs} performs digital processing, digital to analog conversion, analog to digital conversion, power amplification, and filtering, while baseband processing is performed in a \acrshort{cp} connected to multiple \acrshort{bs}s via finite capacity backhaul links. \acrshort{cran} is an emerging network architecture that enables large-scale cooperation among base stations \cite{Peng2016}. The characteristic of this architecture makes \acrshort{cran} capable of dealing with intensive inter-cell interference in future ultra-dense, multi-tier networks \cite{Chen2011}.
The concept of \acrshort{cran} was initially suggested in \cite{Lin2010}. Comprehensive surveys on \acrshort{cran} can be found in  \cite{Checko2015,Simeone2016,Ejaz2020}.

We consider a multi-hop point-to-point communication scheme. The \acrshort{cp} delivers its message over finite capacity noiseless links to three collaborating radio relays (\acrshort{bs}s) that are connected via a \acrfull{mac} to the receiver. This configuration serves as a simple model for a downlink of \acrshort{cran} which is an emerging cellular architecture with centralized processing. Centralized processing schemes facilitate a prominent expansion of the communication bandwidth.

Various representing setups were considered in the information theory literature for the \acrshort{cran} framework. The problem of point-to-point communication over the broadcast channel with the help of two relays connected via finite capacity links to the destination has been addressed in \cite{Sanderovich2008a}. 
A generalized compression strategy for the downlink \acrshort{cran} was proposed in \cite{Patil2019}. The downlink of symmetric \acrshort{cran}s with multiple, non-collaborating relays and a single receiver was studied in \cite{SaeediBidokhti2017}. Lower and upper bounds on the capacity were derived. The lower bound was achieved via Marton's coding, which utilizes dependencies among the various channel inputs. Ozarow's technique was applied to establish the upper bound. The diamond channel consists of an encoder connected via finite capacity fronthaul links to two relays and a decoder. The Gaussian multiple access diamond channel was studied in \cite{Kang2015}. 
The multi-user multi-relay model for the uplink \acrshort{cran} communication with oblivious relays was considered in \cite{EstellaAguerri2019}. In that model, the relay nodes are constrained to operate without knowing the users' codebook. Optimal relay coding schemes were presented, the capacity was determined under the oblivious processing
regime, and connection to the information bottleneck method was settled. Defining the oblivious relay processing region to the downlink \acrshort{cran} is more challenging since information is conveyed first to the relays.
An improved outer bound on the capacity of the downlink \acrshort{cran}, based on the generalized \acrfull{epi}, has been obtained in \cite{Yang2019}.
Capacity approximation within a constant gap of the fronthaul-limited uplink and downlink \acrshort{cran} using noisy network coding and distributed decode-forward has been recently shown in \cite{Ganguly2021}. 
An elegant uplink-downlink duality property for the Gaussian \acrshort{cran} has been identified in \cite{Liu2021}. 

Incorporating inter-relay cooperation has the potential to increase information rates further.  A class of diamond networks with conferencing relays, which is a simple model of \acrshort{cran} with \acrshort{bs}s cooperation, was proposed in \cite{Zhao2015}. The inclusion of cooperation between the \acrshort{bs}s in the downlink \acrshort{cran} model with a pair of mobile users has been addressed in \cite{Wang2018}.  Various coding schemes were proposed and compared. Moreover, \cite{Wang2018} generalizes \cite{Kang2015,SaeediBidokhti2016, Zhao2015} which are shown as special cases.
A similar model, without relay cooperation, has been addressed in \cite{Alqudah2021}.
A communication network consisting of $ k $-transmitters over a \acrfull{mac} with encoder-level cooperation and a single receiver node has been considered in \cite{Noorzad2018}.

In this work, we study point-to-point communication with multiple cooperating relays. Note that this is not a conferencing \acrshort{mac} model \cite{Wigger2008}, but rather has a nontrivial encoding complexity structure, and different coding techniques are employed. Our main contribution is the extension of the two conferencing relays setting \cite{Zhao2015,Wang2018}, and the inclusion of cooperation in the multiple relay setup studied in \cite{SaeediBidokhti2017}. We derive achievability bounds and evaluate them for the \acrfull{awgn} channel. Furthermore, we derive a cooperation dependent upper bound, which is tight in some scenarios. 


%
%

	\section{Problem Formulation}
	\label{section:problem_formulation}
	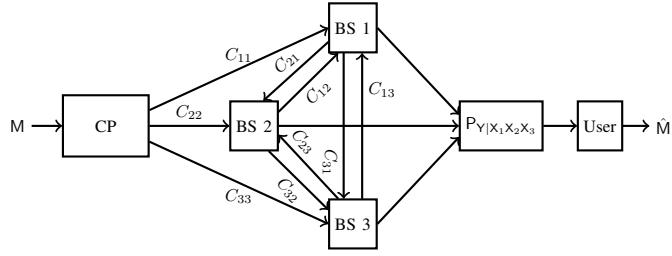
\begin{figure}[t]
	\centering
	\begin{tikzpicture}[thick,scale=0.65, every node/.style={scale=0.65}]
	\coordinate (in) at (-1.5,0);
	\node[draw=black, text width=1.5cm, align = center, minimum height=1.25cm] (cp) at (0,0) {\acrshort{cp}};
	\node[draw=black, align = center, minimum height=1cm] (bs1) at (5,2) {BS 1};
	\node[draw=black, align = center, minimum height=1cm] (bs2) at (3,0) {BS 2};
	\node[draw=black, align = center, minimum height=1cm] (bs3) at (5,-2) {BS 3};
	
	\node[draw=black, align = center, minimum height=1cm] (ch) at (8,0) {$ \mathsf{P}_{\rv{Y}|\rv{X}_1\rv{X}_2\rv{X}_3} $};
	
	\node[draw=black, align = center, minimum height=1cm] (user) at (10,0) {User};
	
	\coordinate (out) at (11,0);
	
	\draw[->] (in) node [left] {$ \rv{M} $}-- (cp);
	
	\draw[->] (cp.20) -- node[above] {$ C_{11} $} (bs1.180);
	\draw[->] (cp.0) -- node[above] {$ C_{22} $} (bs2.180);
	\draw[->] (cp.-20) -- node[below] {$ C_{33} $} (bs3.180);
	
	\draw[->] (bs1.210) -- node[above, sloped] {$ C_{21} $} (bs2.70);
	\draw[->] (bs1.250) -- node[below,sloped, pos=0.75] {$ C_{31} $} (bs3.110);
	
	\draw[->] (bs3.70) -- node[right, pos=0.75] {$ C_{13} $} (bs1.290);
	\draw[->] (bs3.120) -- node[above,sloped, pos=0.75] {$ C_{23} $} (bs2.340);
	
	\draw[->] (bs2.30) -- node[below,sloped] {$ C_{12} $} (bs1.240);
	\draw[->] (bs2.300) -- node[below,sloped] {$ C_{32} $} (bs3.150);
	
	\draw[->] (bs1.0) --  (ch.165);
	\draw[->] (bs2.0) --  (ch.180);
	\draw[->] (bs3.0) --  (ch.195);
	
	\draw[->] (ch) -- (user);
	
	\draw[->] (user) -- (out) node[right] {$ \hat{\rv{M}} $};
	
\end{tikzpicture}
	\caption{Downlink \acrshort{cran} with \acrshort{bs} cooperation: 3 base stations and one mobile user.}
	\label{figure:3bs_1mu_block_diagram}
\end{figure}

Consider the downlink 3-\acrshort{bs} 1-user \acrshort{cran} with \acrshort{bs} cooperation depicted in \autoref{figure:3bs_1mu_block_diagram}. The network consists of one \acrshort{cp}, three \acrshort{bs}s, and one \acrfull{mu}. The \acrshort{cp} communicates with the three BSs through individual noiseless fronthaul links of finite capacities. Denote by $ C_{kk} $ the link's capacity from the \acrshort{cp} to \acrshort{bs} $ k $, for $ k=1,2,3 $. In addition, the three \acrshort{bs}s  can also communicate with each other through individual noiseless fronthaul links of finite capacities. Denote by $ C_{kj} $ the link's capacity from \acrshort{bs} $ j $ to \acrshort{bs} $ k $, where $ j\neq k = 1,2,3 $. The network from the \acrshort{bs}s to the \acrshort{mu} is modeled as a \acrfull{dmmac} $ \langle \set{X}_1 \times \set{X}_2 \times \set{X}_3, \mathsf{P}_{\rv{Y}|\rv{X}_1 \rv{X}_2 \rv{X}_3} , \set{Y} \rangle$ that consists of four finite sets $ \set{X}_1,\set{X}_2,\set{X}_3,\set{Y} $, and a collection of the conditional probability mass functions (\acrshort{pmf}) $ \mathsf{P}_{\rv{Y}|\rv{X}_1 \rv{X}_2 \rv{X}_3} $.

With the help of the three \acrshort{bs}s, the \acrshort{cp} wishes to communicate a message $ \rv{M} $ to the \acrshort{mu}. Assume that $ \rv{M} $ is uniformly distributed over $ \setrn{R}{n} $, where $ \setrn{R}{n} $ is the set of consecutive integers from $ 1 $ to $ 2^{nR} $. This paper restricts
attention to information processing on a block-by-block basis. Each block consists of a sequence of $ n $ symbols. The entire communication is divided into three sequential phases:
\begin{enumerate}
	\item \emph{\acrshort{cp} to \acrshort{bs}s:}
	The \acrshort{cp} conveys three indices $ (\rv{W}_{11},\rv{W}_{22},\rv{W}_{33}) = f_0(\rv{M}) $ to \acrshort{bs} 1, \acrshort{bs} 2, and \acrshort{bs} 3 respectively, where $ f_0 : \setrn{R}{n} \rightarrow \setrn{C_{11}}{n} \times \setrn{C_{22}}{n} \times \setrn{C_{33}}{n} $ is the encoder of the \acrshort{cp}.
	\item \emph{\acrshort{bs}-to-\acrshort{bs} conferencing cooperation:}
	\acrshort{bs} 1 conveys two indices $ (\rv{W}_{21}, \rv{W}_{31}) = f_1(\rv{W}_{11}) $ to \acrshort{bs} 2 and \acrshort{bs} 3, respectively, where $ f_1: \setrn{C_{11}}{n} \rightarrow \setrn{C_{21}}{n} \times \setrn{C_{31}}{n} $ is the conferencing encoder of \acrshort{bs} 1. \acrshort{bs} 2 conveys two indices $ (\rv{W}_{12}, \rv{W}_{32}) = f_2(\rv{W}_{22}) $ to \acrshort{bs} 1 and \acrshort{bs} 3, respectively, where $ f_2: \setrn{C_{22}}{n} \rightarrow \setrn{C_{12}}{n} \times \setrn{C_{32}}{n} $ is the conferencing encoder of \acrshort{bs} 2. Similarly, \acrshort{bs} 3 conveys two indices $ (\rv{W}_{13}, \rv{W}_{23}) = f_3(\rv{W}_{33}) $ to \acrshort{bs} 1 and \acrshort{bs} 2, respectively, where $ f_3: \setrn{C_{33}}{n} \rightarrow \setrn{C_{13}}{n} \times \setrn{C_{23}}{n} $ is the conferencing encoder of \acrshort{bs} 3.
	
	\item \emph{\acrshort{bs}s to the mobile user}:
	\acrshort{bs} $ k $ transmits a sequence $ \rv{X}_k^n = g_k(\rv{W}_{k1},\rv{W}_{k2} ,\rv{W}_{k3}) $ over the \acrshort{dmmac}, where, $ g_k: \setrn{C_{k1}}{n} \times \setrn{C_{k2}}{n} \times \setrn{C_{k3}}{n} \rightarrow \set{X}_k^n $ is the channel encoder of \acrshort{bs} $ k $, $ k\in \{1,2,3\} $.
\end{enumerate}

	Upon receiving the sequence $ \rv{Y}^n $, the \acrshort{mu} assigns an estimate $ \hat{\rv{M}} = d(\rv{Y}^n) $ of the message $ \rv{M} $ where $ d: \set{Y}^n \rightarrow \setrn{R}{n} $. The collection of the encoders $ f_0,f_1,f_2,f_3,g_1,g_2,g_3 $ and the decoder $ d $ constitute a $ (2^{nR},n) $ code.

	The average probability of error is defined as $ P_e^{(n)}  = \Prob{\hat{\rv{M}} \neq \rv{M}}$. A rate $ R $ is said to be achievable if there exists a sequence of $ (2^{nR},n) $ codes such that $ \lim_{n\rightarrow \infty} P_e^{(n)} = 0 $. The capacity $ C $ is the supremum of all achievable rates.

\begin{remark}
    We have a similar assumption as appeared in \cite{Wang2018}, that the conferencing message of some \acrshort{bs} depends solely on the incoming message from the \acrshort{cp}, rather than additional causal dependence on the messages received from the other \acrshort{bs}s, as was assumed in \cite{Zhao2015}. This assumption will facilitate the derivation of a tighter upper bound by establishing a Markov property on the different message sets. Note that the more relaxed assumption has not been used in the derivation of the lower nor the upper bounds in \cite{Zhao2015}, thus, our results are comparable with \cite{Zhao2015}.
\end{remark}

The results we obtain here for the \acrshort{dmmac} can be readily adapted for \acrshort{mac} with continuous input/output alphabets and input costs applying discretization arguments as in \cite[Sec. 3.4.1]{Gamal2011}. More specifically, in this work, we consider the Symmetric Gaussian \acrshort{mac}, which is defined by the following input-output relation:
\begin{equation} \label{eq:cran_3bs_gaussian_channel}
	\rv{Y} =\rv{X}_1 + \rv{X}_2 + \rv{X}_3 + \rv{Z},
\end{equation}
where $ \rv{X}_{k} \in \field{R} $ is the channel input from \acrshort{bs} $ k $, $ \rv{Y} $ is the channel output, 
and $ \rv{Z} \sim \mathcal{N}(0,1) $ is additive noise. In addition, each \acrshort{bs} has to satisfy an average power constraint $ P $, i.e., $ \frac{1}{n} \sum_{i=1}^n \Exp{\rv{X}}_{ki}^2 \leq P $, for all $ k\in \{1,2,3\} $. Furthermore, the noiseless links have symmetric capacities, i.e.,  $ C_{11} = C_{22} = C_{33} = C $, and
$ C_{12} = C_{21} = C_{13} = C_{31} = C_{23} = C_{32}  = C_{0} $.

	\section{The case of 2 relays}

	The downlink 2-\acrshort{bs} 1-user \acrshort{cran}, with \acrshort{bs}s cooperation \cite{Zhao2015} can be considered as a special case of the downlink 3-\acrshort{bs} 1-user \acrshort{cran} defined in \autoref{section:problem_formulation} by setting $ C_{33}=C_{13}=C_{31} = C_{23} = C_{32} = 0 $. 
	\subsection{Bounds on Capacity}
The following lower bound has been obtained in \cite{Wang2018}.
\begin{lemma} [{\cite[Corollary 5]{Wang2018}}] \label{lemma:cran_2bs_dmc_lower_bound}
	Any rate $ R $ is achievable for the downlink 2-\acrshort{bs} 1-user \acrshort{cran} with \acrshort{bs} cooperation if there exists some \acrshort{pmf} $ \mathsf{P}_{\rv{U}\rv{X}_1\rv{X}_2 \rv{Y}} = \mathsf{P}_{\rv{U}\rv{X}_1\rv{X}_2} \mathsf{P}_{\rv{Y}|\rv{X}_1\rv{X}_2}$, $ \rv{U} \in \set{U} $ and $ |\set{U}| \leq \min \{ |\set{X}_1| \cdot |\set{X}_2|+2, |\set{Y}|+4\} $ such that
	\begin{equation*}
		R \mkern-5mu < \mkern-5mu \min \mkern-5mu \left\{ \mkern-14mu
		\begin{array}{ll}
			C_{11}+C_{22} -I(\rv{X}_1;\rv{X}_2|\rv{U}),  \\
			C_{11}+C_{12}+I(\rv{X}_2;\rv{Y}|\rv{U},\rv{X}_1), \\
			C_{22}+C_{21}+I(\rv{X}_1;\rv{Y}|\rv{U},\rv{X}_2) ,\\
			I(\rv{X}_1,\rv{X}_2;\rv{Y}), \\
			\frac{1}{2} \mkern-5mu \left[
				C_{\scaleto{11}{3pt}} \plus C_{\scaleto{22}{3pt}} \plus C_{\scaleto{12}{3pt}} \mkern-5mu +  C_{\scaleto{21}{3pt}} \plus I(\rv{X}_1,\rv{X}_2;\rv{Y}|\rv{U}) \minus I(\rv{X}_1;\rv{X}_2|\rv{U})
			\right] .
		\end{array}
		\right.
	\end{equation*}
\end{lemma}
This bound was initially proposed in \cite{Wang2018} and it recovers the achievability results from \cite[Theorem 2]{Zhao2015} by setting $ C_{12}=C_{21}=C_0 $.
It can be further shown that the proposed scheme also retrieves the achievable rate for the downlink 2-BS 1-user C-RAN without BS cooperation considered in \cite{Kang2015,SaeediBidokhti2016}.

The following computable upper bounds  presented here are  based on the ideas first developed in \cite{SaeediBidokhti2016}.
 \begin{theorem} \label{theorem:cran_2bs_upper_bound1}
 	An upper bound on the capacity of the diamond channel with two conferencing  relays is given by 
 		$ C < \max_{\mathsf{P}_{\rv{X}_1 \rv{X}_2}} \min_{\mathsf{P}_{\rv{V}|\rv{U}\rv{X}_1 \rv{X}_2 \rv{Y}}} \max_{\mathsf{P}_{\rv{U}|\rv{X}_1 \rv{X}_2} } \beta_1 $,
 	where
 	\begin{equation} \label{eq:upper_bound_2relays_noQ}
 		\beta_1 \triangleq  \min 
 		\left\{
 		\begin{aligned}
 			& C_{11}+C_{22}, \\
 			& C_{11}+ C_{12}  + I(\rv{X}_{2};\rv{Y}|\rv{U},\rv{X}_{1}), \\
 			& C_{22}+ C_{21}  + I(\rv{X}_{1};\rv{Y}|\rv{U},\rv{X}_{2}), \\
 			& I(\rv{X}_1,\rv{X}_2;\rv{Y}), \\
 			& C_{12}+C_{21} + I(\rv{X}_1,\rv{X}_2;\rv{Y}|\rv{U}), \\
 			&\frac{1}{2} \left[ 
 			\begin{aligned}
 				&C_{11}+C_{2} + C_{12} + C_{21} + I(\rv{X}_1;\rv{V}|\rv{U},\rv{X}_2)\\ 
 				&
 				 + I(\rv{X}_{1},\rv{X}_{2};\rv{Y}|\rv{U},\rv{V})  +I(\rv{X}_2;\rv{V}|\rv{U},\rv{X}_1) 
 			\end{aligned}
 			  \right].
 		\end{aligned}
 		\right.
 	\end{equation}
 \end{theorem}
The proof is postponed to Appendix \ref{appendix:cran_2bs_upper_bound1_proof}. 
\begin{remark} \label{remark:upper_bound_equivalent_form}
	Choosing  $ \mathsf{P}_{\rv{V}|\rv{U},\rv{X}_1,\rv{X}_2,\rv{Y}} = \mathsf{P}_{\rv{V}|\rv{Y}} $, the last term in the brackets of \eqref{eq:upper_bound_2relays_noQ} can be equivalently rewritten as 
	\begin{equation} \label{eq:upper_bound_equivalent_term}
		C \leq \frac{1}{2} \left[
		\begin{aligned}
			 C_{11}+C_{22} + C_{12} + C_{21}  +  I(\rv{X}_{1},\rv{X}_{2};\rv{Y}|\rv{U})  \\
			+ I(\rv{X}_1;\rv{X}_2|\rv{U},\rv{V})
			- I(\rv{X}_1;\rv{X_2}|\rv{U}) 
		\end{aligned}
		  \right].
	\end{equation}
	We will exploit this representation to design $\rv{V}$ that minimizes $I(\rv{X}_1;\rv{X}_2|\rv{U},\rv{V})$.
\end{remark}

\begin{remark}
	The bound in \autoref{theorem:cran_2bs_upper_bound1} is based on \cite[Thm. 2]{SaeediBidokhti2016}. Note that it coincides with \cite[Thm. 1]{Zhao2015} by setting $\set{U} = \emptyset$, but, the last term in \eqref{eq:upper_bound_equivalent_term} suggests it may not be the optimal choice. Thus, this bound is a promising candidate.
\end{remark}

Our next upper bound is  based on \cite[Thm. 3]{SaeediBidokhti2016}. This bound also incorporates the collaborative nature of the problem and therefore is tighter then the respective cut-set bound.

\begin{theorem}\label{theorem:cran_2bs_upper_bound2}
	An upper bound on the capacity of 2-\acrshort{bs}s 1-user network with conferencing relays is given by
		\[C < \max_{\mathsf{P}_{\rv{X}_{1} \rv{X}_2}} \min_{\mathsf{P}_{\rv{V}|\rv{Y}}} \max_{\mathsf{P}_{\rv{U} \rv{T}|\rv{X}_1 \rv{X}_2} } \beta_2, \]
	where
	\begin{equation} \label{eq:upper_bound2x}
		  \beta_2 = \min
		\left\{
		\begin{aligned}
			&C_{11}+C_{22},\\
			&C_{11}+C_{12}+I(\rv{X}_2;\rv{Y}|\rv{U},\rv{X}_1,\rv{T}), \\
			&C_{22}+C_{21}+I(\rv{X}_1;\rv{Y}|\rv{U},\rv{X}_2,\rv{T}), \\
			&I(\rv{X}_1,\rv{X}_2;\rv{Y}), \\
			&C_{12}+C_{21}+I(\rv{X}_1,\rv{X}_2;\rv{Y}|\rv{U},\rv{T}),\\
			& \begin{aligned}
				&C_{11}+C_{22} -I(\rv{X}_1,\rv{X}_2;\rv{V}|\rv{U},\rv{T}) + \\ &I(\rv{X}_1;\rv{V}|\rv{U},\rv{X}_2,\rv{T}) + I(\rv{X}_2;\rv{V}|\rv{U},\rv{X}_1,\rv{T}).
			\end{aligned}
		\end{aligned}
		\right\}
	\end{equation}
\end{theorem}
The complete proof is postponed to Appendix \ref{appendix:cran_2bs_upper_bound2_proof}.

	\subsection{Example: Gaussian \acrshort{mac}}
	In this section we  provide upper and lower bounds on the capacity of the  2-BS 1-user C-RAN with a symmetric Gaussian \acrshort{mac}, defined by \eqref{eq:cran_3bs_gaussian_channel} when $\rv{X}_3= 0$.
Plugging jointly Gaussian $ (\rv{U},\rv{X}_1,\rv{X}_2) $ in \autoref{lemma:cran_2bs_dmc_lower_bound} we obtain the following achievable rate.
\begin{proposition} \label{proposition:cran_2bs_gaussian_lower_bound}
	For the Symmetric Gaussian diamond \acrshort{mac} with two conferencing relays, rate $ R $ is achievable if for some $ 0\leq \rho \leq 1 $, $ 0\leq \rho_c \leq \sqrt{\frac{1+\rho}{2}} $, it satisfies
	\begin{equation*}
		R \leq \min 
		\begin{cases}
			2C -\frac{1}{2} \log \left(\frac{\left(1-\rho_c^2\right)^2}{(1-\rho)(1+\rho -2\rho_c^2)}\right) ,\\
			C+C_{0} +\frac{1}{2} \log \left(\frac{1- \rho_c^2 +  (1-\rho) (1+ \rho - 2\rho_c^2) P}{1-
				\rho_c^2}\right), \\
			\frac{1}{2} \log \left(1 +2(1+ \rho) P \right) ,\\
			C \mkern-5mu + \mkern-5mu C_0  \mkern-5mu + \mkern-5mu \frac{1}{4} \mkern-5mu  \log \mkern-5mu \left( \frac{ [1    +    2(1     +  \rho \minus 2\rho_c^2) P ](1-\rho)(1+\rho -2\rho_c^2)}{\left(1-\rho_c^2\right)^2} \right)  \mkern-5mu .
		\end{cases}
	\end{equation*}
\end{proposition}
The proof is postponed to Appendix \ref{appendix:cran_2bs_gaussian_lower_bound_proof}. 

We proceed to derive upper bounds for the symmetric case. We choose $ \rv{V} $ to be a noisy version of $ \rv{Y} $, i.e., $ \rv{U} = \rv{Y} + \rv{W}  $ where $ \rv{W} \sim (0,N) $.  The following upper bound is a specialization of \autoref{theorem:cran_2bs_upper_bound1} to a Gaussian setting, utilizing \autoref{remark:upper_bound_equivalent_form} and choosing $ N $ such that $ I(\rv{X}_1;\rv{X}_2|\rv{U},\rv{V}) $ is  zero if possible.
\begin{proposition}
	\label{proposition:cran_2bs_gaussian_upper_bound1}
	Rate $ R $ is achievable for the Symmetric Gaussian Diamond \acrshort{mac} only if it satisfies the following constraints for some $ 0\leq  \rho_c \leq \sqrt{\frac{1+\rho}{2}} $, $ 0 \leq \rho \leq \rho^* $ :
	\begin{equation} \label{eq:cran_gaussian_sym_ub_rholessrhostar}
		R \leq \min 
		\begin{cases}
			2C,\\
			C+C_{0} +\frac{1}{2} \log \left(\frac{1- \rho_c^2 +  (1-\rho) (1+ \rho - 2\rho_c^2) P}{1-
				\rho_c^2}\right), \\
			\frac{1}{2} \log \left(1 +2(1+ \rho) P \right), \\
			2C_{0}+\frac{1}{2} \log \left(  1+2(1 + \rho - 2\rho_c^2 ) P \right),\\
			C \mkern-5mu + \mkern-5mu C_0  \mkern-5mu + \mkern-5mu \frac{1}{4} \mkern-5mu  \log \mkern-5mu \left( \frac{ [1    +    2(1     +  \rho \minus 2\rho_c^2) P ](1-\rho)(1+\rho -2\rho_c^2)}{\left(1-\rho_c^2\right)^2} \right)  \mkern-5mu ,
		\end{cases}
	\end{equation}
	and for $ \rho > \rho^* $ we have
	\begin{equation} \label{eq:cran_gaussian_sym_ub_rhogreaterrhostar}
		R \leq \min 
		\begin{cases}
			2C,\\
			C+C_{0} +\frac{1}{2} \log \left(\frac{1- \rho_c^2 +  (1-\rho) (1+ \rho - 2\rho_c^2) P}{1-
				\rho_c^2}\right), \\
			\frac{1}{2} \log \left(1 +2(1+ \rho) P \right), \\
			2C_{0}+\frac{1}{2} \log \left(  1+2(1 + \rho - 2\rho_c^2 ) P \right),
		\end{cases}
	\end{equation}
	where 
	$
		\rho^* \triangleq \rho_c^2 + \sqrt{1 + \frac{1}{4 P^2} - 2\rho_c^2 + \rho_c ^4} - \frac{1}{2P}
	$.
\end{proposition}
\begin{remark}
	Note that if $ \rho > \rho^* $, then the maximum of every term in \eqref{eq:cran_gaussian_sym_ub_rhogreaterrhostar} is achieved with $ \rho_c = 0 $. In such case, the last term becomes degenerated.
	Thus, the interesting regime is when $ \rho < \rho^* $ and the optimal $ \rho_c > 0 $. The only term that may contribute to this outcome is the last term in \eqref{eq:cran_gaussian_sym_ub_rholessrhostar}.
	The problem is that term is coupled with 
	\begin{equation*}
		C+C_{0} +\frac{1}{2} \log \left(\frac{1- \rho_c^2 +  (1-\rho) (1+ \rho - 2\rho_c^2) P}{1-
			\rho_c^2}\right),
	\end{equation*}
	which is maximized with $ \rho_c = 0 $ for any fixed $ \rho $.
\end{remark}

Finally, we specialize \autoref{theorem:cran_2bs_upper_bound2} for the Gaussian setting.
\begin{proposition}
		\label{proposition:cran_2bs_gaussian_upper_bound2}
	Rate $ R $ is achievable for the Symmetric Gaussian Diamond \acrshort{mac} only if it satisfies the following constraints for some $ 0\leq  \rho_c \leq \sqrt{\frac{1+\rho}{2}} $, $ 0 \leq \rho \leq 1 $ ,
	$		R \leq \min_{N\geq 0}  \min \beta_3 $,
	where
	\begin{equation*}
		\beta_3 \mkern-7mu = \mkern-6mu
		\begin{cases} \mkern-5mu
			2C, \\
			C+C_{0} +\frac{1}{2} \log \left(\frac{1- \rho_c^2 +  (1-\rho) (1+ \rho - 2\rho_c^2) P}{1-
				\rho_c^2}\right), \\
			\frac{1}{2} \log \left(1 +2(1+ \rho) P \right), \\
			2C_{0}+\frac{1}{2} \log \left(  1+2(1 + \rho - 2\rho_c^2 ) P \right), \\
			\mkern-5mu \frac{1}{2} \mkern-5mu \log \mkern-5mu \left[ \mkern-8mu
			\sqrt{
				 \mkern-3mu   \frac{ 2^{4(C+C_0)}
					[\bar{\rho}_c ^2(1+N) + \bar{\rho}(1+\rho \minus 2\rho_c^2)P]^2}{
					(1-\rho_c^2)^2(1+N)
				} \mkern-5mu + \mkern-5mu \frac{2^{8 C_{0}} N^2}{4} } \mkern-2mu \minus \mkern-2mu \frac{2^{4 C_{0}} N }{2} \mkern-5mu
			\right] \mkern-5mu.
		\end{cases}
	\end{equation*}
\end{proposition}

\begin{figure}
	\centering
	\input{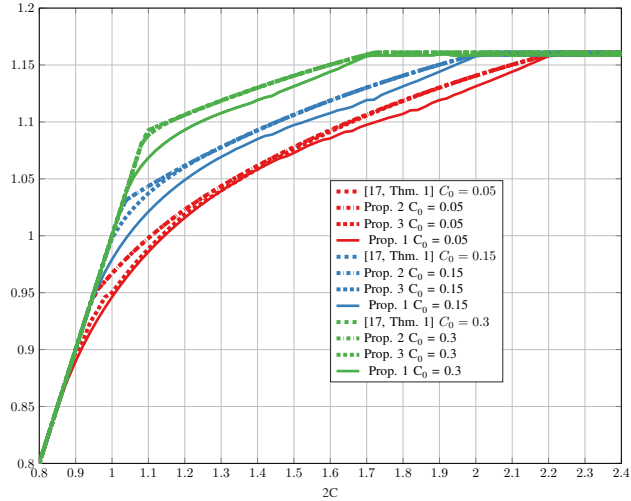}
	\caption{Lower and upper bounds for 2 relays scenario from Props. \ref{proposition:cran_2bs_gaussian_lower_bound}, \ref{proposition:cran_2bs_gaussian_upper_bound1} and \ref{proposition:cran_2bs_gaussian_upper_bound2}, with various values of $ C_0 $ and $ P=1 $ vs $ C $.}
	\label{figure:cran_2bs_bounds_p_1}
\end{figure}

We further give a representative evaluation of the proposed bounds from Props. \ref{proposition:cran_2bs_gaussian_lower_bound}, \ref{proposition:cran_2bs_gaussian_upper_bound1} and \ref{proposition:cran_2bs_gaussian_upper_bound2} for $ P=1 $  in \autoref{figure:cran_2bs_bounds_p_1}.

The proof of the upper bounds from Props. \ref{proposition:cran_2bs_gaussian_upper_bound1} and \ref{proposition:cran_2bs_gaussian_upper_bound2}, is postponed to Appendix \ref{appendix:cran_2bs_gaussian_upper_bounds_proof}.

	\section{The case of 3 relays - Symmetric}
	\subsection{Lower Bounds}
\begin{figure}[t]
	\centering
	\begin{tikzpicture}[thick,scale=0.825, every node/.style={scale=0.825}]
	\def\y{1.2}
	\def\yy{2.4}
	
	\def\yyy{3.6}
	
	\draw (0,0) -- (1,0);
	\draw (0,0.1) -- (1,0.1);
	\draw (0,0.4) -- (1,0.4);
	\draw (0,0.5) -- (1,0.5);
	\draw (0,0.6) -- (1,0.6);
	\draw (0,0.7) -- (1,0.7);
	\draw (0,0.8) -- (1,0.8);
	\draw (0,0.9) -- (1,0.9);
	\draw (0,1) -- (1,1);
	
	\draw (0,0) -- (0,1);
	\draw (1,0) -- (1,1);
	
	\draw (0,\y) -- (1,\y);
	\draw (0,\y+0.1) -- (1,\y+0.1);
	\draw (0,\y+0.4) -- (1,\y+0.4);
	\draw (0,\y+0.5) -- (1,\y+0.5);
	\draw (0,\y+0.6) -- (1,\y+0.6);
	\draw (0,\y+0.7) -- (1,\y+0.7);
	\draw (0,\y+0.8) -- (1,\y+0.8);
	\draw (0,\y+0.9) -- (1,\y+0.9);
	\draw (0,\y+1) -- (1,\y+1);
	
	\draw (0,\y) -- (0,\y+1);
	\draw (1,\y) -- (1,\y+1);
	
	\draw (0,\yy) -- (1,\yy);
	\draw (0,\yy+0.1) -- (1,\yy+0.1);
	\draw (0,\yy+0.4) -- (1,\yy+0.4);
	\draw (0,\yy+0.5) -- (1,\yy+0.5);
	\draw (0,\yy+0.6) -- (1,\yy+0.6);
	\draw (0,\yy+0.7) -- (1,\yy+0.7);
	\draw (0,\yy+0.8) -- (1,\yy+0.8);
	\draw (0,\yy+0.9) -- (1,\yy+0.9);
	\draw (0,\yy+1) -- (1,\yy+1);
	
	\draw (0,\yy) -- (0,\yy+1);
	\draw (1,\yy) -- (1,\yy+1);
	
	\draw (0,\yyy) -- (1,\yyy);
	\draw (0,\yyy+0.1) -- (1,\yyy+0.1);
	\draw (0,\yyy+0.4) -- (1,\yyy+0.4);
	\draw (0,\yyy+0.5) -- (1,\yyy+0.5);
	\draw (0,\yyy+0.6) -- (1,\yyy+0.6);
	\draw (0,\yyy+0.7) -- (1,\yyy+0.7);
	\draw (0,\yyy+0.8) -- (1,\yyy+0.8);
	\draw (0,\yyy+0.9) -- (1,\yyy+0.9);
	\draw (0,\yyy+1) -- (1,\yyy+1);
	
	\draw (0,\yyy) -- (0,\yyy+1);
	\draw (1,\yyy) -- (1,\yyy+1);
	
	\node (u0) at (-0.3,\yyy+0.4) {$ \rv{U}_0 $};
	\node (u1) at (-0.3,\yy+0.4) {$ \rv{U}_1 $};
	\node (u2) at (-0.3,\y+0.4) {$ \rv{U}_2 $};
	\node (u3) at (-0.3,0.4) {$ \rv{U}_3 $};

	\def\by{3.7}
	\def\byy{2.9}
	\def\byyy{2.1}
	\def\byyyy{1.3}
	\def\byyyyy{0.5}
	
	\draw[fill=gray, opacity=0.6] (2,\by) -- (2.5,\by) -- (2.5,\by+0.8) -- (2,\by+0.8) -- cycle;
	\draw (2,\byy) -- (2.5,\byy) -- (2.5,\byy+0.8) -- (2,\byy+0.8) -- cycle;
	\draw (2,\byyy) -- (2.5,\byyy) -- (2.5,\byyy+0.8) -- (2,\byyy+0.8) -- cycle;
	\draw (2,\byyyy) -- (2.5,\byyyy) -- (2.5,\byyyy+0.8) -- (2,\byyyy+0.8) -- cycle;
	\draw (2,\byyyyy) -- (2.5,\byyyyy) -- (2.5,\byyyyy+0.8) -- (2,\byyyyy+0.8) -- cycle;
	
	\node (bm) at (2.25,\by+1) {$ \set{B}(m) $};

	\node[circle,draw=black,  inner sep=0pt,minimum size=3pt] (b11) at (2.35,\by+0.6) {};
	\node[circle,draw=black,  inner sep=0pt,minimum size=3pt] (b12) at (2.35,\by+0.3) {};
	\node[circle,draw=black,  inner sep=0pt,minimum size=3pt] (b13) at (2.15,\by+0.4) {};
	\node[circle,draw=black,  inner sep=0pt,minimum size=3pt] (b14) at (2.15,\by+0.2) {};
	
	\node[circle,draw=black,  inner sep=0pt,minimum size=3pt] (b21) at (2.35,\byy+0.6) {};
	\node[circle,draw=black,  inner sep=0pt,minimum size=3pt] (b22) at (2.35,\byy+0.3) {};
	\node[circle,draw=black,  inner sep=0pt,minimum size=3pt] (b23) at (2.15,\byy+0.4) {};
	\node[circle,draw=black,  inner sep=0pt,minimum size=3pt] (b24) at (2.15,\byy+0.2) {};
	
	\node[circle,draw=black,  inner sep=0pt,minimum size=3pt] (b31) at (2.35,\byyy+0.6) {};
	\node[circle,draw=black,  inner sep=0pt,minimum size=3pt] (b32) at (2.35,\byyy+0.3) {};
	\node[circle,draw=black,  inner sep=0pt,minimum size=3pt] (b33) at (2.15,\byyy+0.4) {};
	\node[circle,draw=black,  inner sep=0pt,minimum size=3pt] (b34) at (2.15,\byyy+0.2) {};

	\node[circle,draw=black,  inner sep=0pt,minimum size=3pt] (b51) at (2.35,\byyyyy+0.6) {};
	\node[circle,draw=black,  inner sep=0pt,minimum size=3pt] (b52) at (2.35,\byyyyy+0.3) {};
	\node[circle,draw=black,  inner sep=0pt,minimum size=3pt] (b53) at (2.15,\byyyyy+0.4) {};
	\node[circle,draw=black,  inner sep=0pt,minimum size=3pt] (b54) at (2.15,\byyyyy+0.2) {};
	
	\draw (1,\yyy+0.5) -- node[above]{$ k_0 $} (b14);
	\draw (1,\yy+0.5) -- node[above,sloped]{$ k_1 $} (b14);
	\draw (1,\y+0.5) -- node[above,sloped]{$ k_2 $} (b14);
	\draw (1,0.5) -- node[below,sloped,xshift = -0.5cm]{$ k_3 $} (b14);
	
	\node[draw=black] (typ) at (4.5, 2.5) {$ \typsetn{\epsilon'}{\rv{U}_{\Omega}} $};
	
	\draw (2.5,\by+0.8) -- (typ.180);
	\draw (2.5,\by) -- (typ.180);
	
	\node[draw=black] (d1) at (7, 0) {$ \set{D}_1 $};
	\node[draw=black] (d2) at (7, 1.5) {$ \set{D}_2 $};
	\node[draw=black] (d0) at (7, 3) {$ \set{D}_0 $};
	\node[draw=black] (d3) at (7, 4.5) {$ \set{D}_3 $};
	
	\draw (typ.0) -- node[above,sloped] {\tiny{$ k_0 $}} (d0);
	\draw (typ.0) -- node[below,sloped] {\tiny{$ (k_0,k_1) $}} (d1);
	\draw (typ.0) -- node[above,sloped] {\tiny{$ (k_0,k_2) $}} (d2);
	\draw (typ.0) -- node[above,sloped] {\tiny{$ (k_0,k_3) $}} (d3);
	
	\node[draw=black,minimum height=1cm] (bs1) at (9.5, 0) {BS 1};
	\node[draw=black,minimum height=1cm] (bs2) at (9.5, 1.5) {BS 2};
	\node[draw=black,minimum height=1cm] (bs3) at (9.5, 4.2) {BS 3};
	
	\draw[->] (d0.10) -- (d0.10-|8.9,1) node[right]{\tiny{$ m_{00} $}} --    (8.9,1|-bs3.220) -- (bs3.220);
	\draw[->] (d0.10-|8.9,1) -- (8.9,1|-bs2.140) -- (bs2.140);
	\draw[->] (8.9,1|-bs2.140) -- (8.9,1|-bs1.140)-- (bs1.140);
	
	\draw[->] (d0.30) -- node[above=-2pt,sloped]{\tiny{$ m_{03} $}}(d0.30-|8.8,1) -- (8.8,1|-bs3.205) -- (bs3.205);
	\draw[->] (d0.-10) -- (d0.-10-|8.8,1) -- node[below=-2pt,sloped]{\tiny{$ m_{02} $}} (8.8,1|-bs2.150) -- (bs2.150);	
	\draw[->] (d0.-30) -- node[below=-2pt,sloped]{\tiny{$ m_{01} $}}(d0.-30-|8.5,1) -- (8.5,1|-bs1.150) -- (bs1.150);	
	
	\draw[->,red] (d1.-20)  -- node[below]{\tiny{$ m_{11} $}} (d1.-20-|bs1.180);
	\draw[->,red] (d1.0)  -- (d1.0-|8,1) -- node[above=-2pt,sloped]{\tiny{$ m_{12} $}}  (8,1|-bs2.200) -- (bs2.200); 
	\draw[->,red] (d1.20)  -- node[above=-2pt,sloped]{\tiny{$ m_{13} $}} (d1.20-|7.75,1) --   (7.75,1|-bs3.190) -- (bs3.190); 
	
	\draw[->,black!60!green] (d2.-20)  --  (d2.-20-|8.2,0.9) -- node[below=-2pt,sloped]{\tiny{$ m_{21} $}} (8.2,0.9|- bs1.180) -- (bs1.180);
	\draw[->,black!60!green] (d2.0)  -- node[above=-2pt,sloped]{\tiny{$ m_{22} $}}  (bs2.180); 
	\draw[->,black!60!green] (d2.20)  --(d2.20-|7.5,1) --  node[above=-2pt,sloped,pos=0.8]{\tiny{$ m_{23} $}}  (7.5,1|-bs3.170) -- (bs3.170);
	
	\draw[->,blue] (d3.-20)  --  (d3.-20-|8.3,1) --  node[below=-2pt,sloped]{\tiny{$ m_{31} $}}  (8.3,1|- bs1.160) -- (bs1.160);
	\draw[->,blue] (d3.0)  -- (d3.0-|8.4,1) --  node[above=-2pt,sloped,pos=0.25]{\tiny{$ m_{32} $}}  (8.4,1|-bs2.160) --   (bs2.160); 
	\draw[->,blue] (d3.20)  --   node[above=-2pt,sloped]{\tiny{$ m_{33} $}}  (d3.20-|bs3.180);
		
\end{tikzpicture}
	\caption{Illustration of the encoding operation at the central processor in the SCC scheme.}
	\label{figure:3bs_1ms_scc_encoding}
\end{figure}
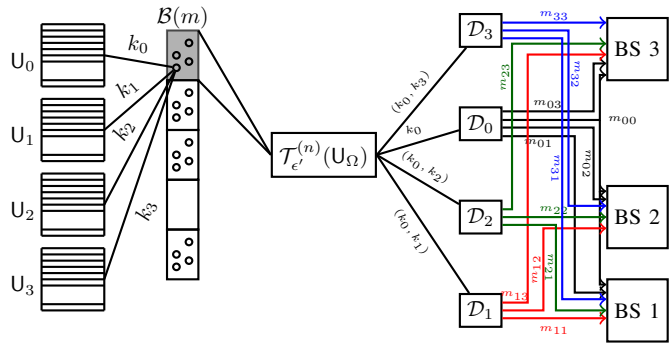

	First, let us give a high-level overview of the proposed  coding strategy. Consider the  set $ \Omega = \{0,1,2,3\} $, fix a joint \acrshort{pmf} $ P_{\rv{U}_{\Omega}} $ and independently generate 4 codebooks with sizes $ \setrn{R_{\omega}}{n} $ from the marginals $ P_{\rv{U}_{\omega}} $, $ \forall \omega \in \Omega $. Each message $ m \in \setrn{R}{n} $ is associated with a unique bin $ \set{B}(m) $ of index tuples $ k_{\Omega} \triangleq (k_0,k_1,k_2,k_3) $, which are indices of the corresponding dictionaries. Then, given $ m $, we apply joint typicality encoding to find index tuples $ k_{\Omega}  \in \set{B}(m) $, such that $ (\rv{U}_\Omega^n(k_{\Omega})) $ are jointly typical. Subsequently, those index tuples are then sent to the respective \acrshort{bs}s. Here, the fact that cooperation exists between the \acrshort{bs}s is used to increase the effective rate from the \acrshort{cp} to \acrshort{bs}s, since there are multiple paths to transmit information to each one of the \acrshort{bs}s. Finally, the \acrshort{mu} applies joint typicality decoding to recover $ k_{\Omega} $ and then the message $ m $. The encoding architecture is illustrated in \autoref{figure:3bs_1ms_scc_encoding}. The resulting achievability rate is presented in the following theorem.

\begin{theorem} \label{theorem:single_common_code}
	Let $ \Omega = \{0,1,2,3\} $. A rate $ R $ is achievable for the downlink 3-\acrshort{bs} 1-user \acrshort{cran} with \acrshort{bs} cooperation if there exist some rates $ R_{\omega} \geq 0 $, $ \omega \in \Omega  $, some joint \acrshort{pmf} $ \mathsf{P}_{\rv{U}_{\Omega}} $ and some functions $ x_1(u_0,u_1) $, $ x_2(u_0,u_{2}) $, $ x_3(u_0,u_3) $, such that for all $ \set{S} \subseteq \Omega  $ satisfying $ |\set{S}| \geq 1 $, the following rate constraints hold:
	\begin{itemize}
		\item 
		$ \indicator \{\set{S} = \Omega\} R  < \sum_{\omega \in \set{S}} R_{\omega} - \Gamma(\rv{U}_{\set{S}}) $;
	\item 
		$ \sum_{\omega \in \set{S}} R_{\omega} < I(\rv{U}_{\set{S}}; \rv{U}_{\set{S}^c},\rv{Y}) + \Gamma (\rv{U}_{\set{S}}) $;
	\item 
		$ \sum_{l \in \Omega} C_{kl}   \geq R_0 + R_k - I(\rv{U}_0;\rv{U}_k)  \quad \forall k \in \Omega $;
\item 		$ \sum_{k \in \Omega} C_{kk}  \geq R_0 + \sum_{k \in \Omega} R_k \minus I(\rv{U}_0;\rv{U}_k)  $.
\end{itemize}
where $\Gamma(\rv{U}_{\set{S}}) \triangleq \sum_{\omega \in \set{S}} H(\rv{U}_{\omega}) - H(\rv{U}_{\set{S}}) $.
\end{theorem}

The proof of this theorem is postponed to Appendix \ref{appendix:cran_3bs_coding_scheme}.
Evaluation of the above rate  for a specific channel is very difficult. Thus, we consider the following corollary where we restrict the correlation structure to be $ \rv{U}_0 = \rv{U}$ and $ \rv{U}_k = \rv{X}_k $ for $ k \in [3] $.
\begin{corollary} \label{corollary:computable_lower_bound}
	Rate $ R $ is achievable for the downlink 3-\acrshort{bs} 1-user \acrshort{cran} with cooperation if there exists $ \mathsf{P}_{\rv{U}\rv{X}_1\rv{X}_2\rv{X}_3} $ such that the following holds for any $ \set{S} \subset [3] $:
	\begin{equation} \label{eq:cran3_common_cloude_rate}
		R \leq \min \left\{
		\begin{aligned}
			&\sum_{ \omega \in [3]} C_{\omega \omega} - \Gamma(\rv{X}_{\Omega}|\rv{U}),  \\
			& \mkern-7mu \sum_{k \in \set{S},\omega \in [3]} \mkern-15mu C_{k \omega} \plus I(\rv{X}_{\set{S}^c};\rv{Y}|\rv{U},\rv{X}_{\set{S}}) \minus \Gamma(\rv{X}_{\set{S}}|\rv{U}), |\set{S}| \geq 1,  \\
			& I(\rv{X}_{[3]};\rv{Y}), \\
			& \frac{1}{2} \left[
			\begin{aligned}
				\sum_{k \in \set{S},\omega \in [3]} C_{k \omega} + I(\rv{X}_{\set{S}^c};\rv{Y}|\rv{U},\rv{X}_{\set{S}})  \\
				+ 				 I(\rv{X}_{[3]};\rv{Y}|\rv{U}) \minus  \Gamma(\rv{X}_{\set{S}}|\rv{U})  
			\end{aligned}
			\right], |\set{S}| =2 \\
			& \frac{1}{3} \left[\sum_{\omega \omega'\in [3]} C_{\omega \omega'} + 2 I(\rv{X}_{[3]};\rv{Y}|\rv{U}) - \Gamma(\rv{X}_{[3]}|\rv{U})  \right].
		\end{aligned}
		\right.
	\end{equation}
\end{corollary}

\begin{remark}
	In our coding scheme for the three relays setting of  \autoref{corollary:computable_lower_bound}, we have an implicit symmetry assumption, i.e., fixing the rates corresponding to the third node to zero will result in congestion of the common rate. Therefore, \autoref{corollary:computable_lower_bound} cannot be directly related to the two relays scenario. Nevertheless, the lines of the equations in the lower bound for two relays of  \autoref{lemma:cran_2bs_dmc_lower_bound} are comparable to the lines of the lower bound for three relays  \autoref{corollary:computable_lower_bound}, except the last line of \eqref{eq:cran3_common_cloude_rate}, reflecting the similarities among the coding schemes. 
\end{remark}

	\subsection{Upper Bound}
	
\begin{figure}[b]
	\centering
	\input{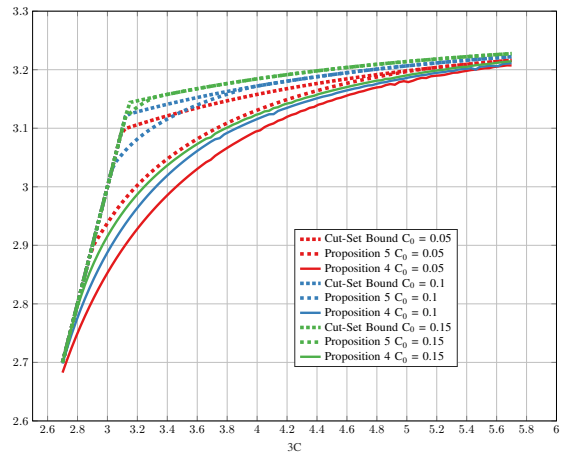}
	\caption{Lower and upper bounds of the 3 relay setting from Props. \ref{proposition:cran_3bs_gaussian_lower_bound} and \ref{proposition:cran_3bs_upper_bound2_gaussian}, for various values of $ C_0 $ and $ P=1 $ vs $ C $.}
	\label{figure:cran_3bs_bounds_p_1}
\end{figure}

We present here a new upper bound on capacity that extends the bounds found in \cite{Zhao2015,SaeediBidokhti2016} for the three relays scenario with cooperation.
\begin{theorem}\label{theorem:cran_3bs_upper_bound2}
	An upper bound on the capacity of 3-\acrshort{bs}s 1-\acrshort{mu} network with conferencing relays is given by
\[ 		R \leq \max_{\mathsf{P}_{\rv{X}_{1} \rv{X}_2}} \min_{\mathsf{P}_{\rv{V}|\rv{Y}}} \max_{\mathsf{P}_{\rv{U} \rv{T}|\rv{X}_1 \rv{X}_2} } \beta_4, \]
	where
	\begin{equation} \label{eq:upper_bound2_3relays_noQ}
		\beta_4 \mkern -2mu =  \mkern -2mu   \min  \mkern -2mu 
		\left\{ \mkern -5mu 
		\begin{aligned}
			& \sum_{ \omega \in \set{S}} C_{\omega \omega}, \\
			&\sum_{ \omega \in \set{S}} C_{\omega \omega} +   \sum_{
				\substack{\omega \in [3] \\ \omega' \neq \omega \in \set{S}^c}
			} C_{\omega\omega'}  +   I(\rv{X}_{\set{S}^c};\rv{Y}|\rv{U},\rv{X}_{\set{S}},\rv{T}),\\
			& I(\rv{X}_{[3]};\rv{Y}) , \\
			& \mkern-5mu \sum_{\omega \in [3]}  \mkern-5mu C_{\omega \omega} \minus  2I(\rv{X}_{[3]};\rv{V}|\rv{U},\rv{T})  \mkern-5mu +  \mkern-5mu I(\rv{X}_{1}, \rv{X}_{2};\rv{V}|\rv{U},\rv{T},\rv{X}_{3}) \\
			&+ I(\rv{X}_{1}, \rv{X}_{3};\rv{V}|\rv{U},\rv{T},\rv{X}_{2})+ I(\rv{X}_{2}, \rv{X}_{3};\rv{V}|\rv{U},\rv{T},\rv{X}_{1}).
		\end{aligned}
		 \mkern -5mu 
		\right.
	\end{equation}
\end{theorem}
This result follows by combining 
methods used to single-letterize the expressions in the proof of \cite[Thm. 3]{SaeediBidokhti2016}. 
The complete proof is postponed to Appendix \ref{appendix:upper_bound2_proof}.

	\subsection{Gaussian \acrshort{mac}}
	In this section we evaluate \autoref{corollary:computable_lower_bound} and \autoref{theorem:cran_3bs_upper_bound2} for the Symmetric Gaussian \acrshort{mac}. 
We define the following functions:
\begin{align*}
	F_1
	&=
	 \frac{1}{2} \log \frac{(1-\rho_c^2)^3}{(1-\rho)^2(1+2\rho-3\rho_c^2)} \\
	 F_2 
	 &= \frac{1}{2} \log \left(\frac{1- \rho_c^2 + 2 \left(1-2\rho ^2-3\rho_c^2+\rho+3  \rho  \rho_c^2\right) P}{1-
	 	\rho_c^2}\right) \\
	 F_3
	 &= \frac{1}{2} \log \left( \mkern-5mu \frac{1+\rho-2\rho_c^2+ \left(1 + \rho-2\rho ^2-3\rho_c^2-3 \rho\rho_c^2 \right) P}{1+\rho-2\rho_c^2
	 } \mkern-5mu \right)  \\
	 F_4
	 &=
 	\frac{1}{2} \log \frac{(1-\rho_c^2)^2}{(1+\rho)(1-\rho +2\rho_c^2)}, \\
 	F_5
 	&= \frac{1}{2} \log [1 \plus 3(1 \plus 2\rho)P]  \\
 	F_6 
 	&= \frac{1}{2} \log \left(  1+3(1 + 2\rho - 3\rho_c^2) P \right) \\
 	F_7
 	&=  -  \log \left(2^{2(R-6C_{0})} + N \right) - \frac{1}{2} \log (1+N)
 	\\
 	&+ 
 	\frac{3}{2} \log 
 	\left(
 	\frac{
 		(1-\rho_c^2)(1+N)+2(1-\rho)(1+2\rho-3\rho_c^2)P}{
 		1-\rho_c^2	
 	}
 	\right) .
\end{align*}


Utilizing \autoref{corollary:computable_lower_bound} for the symmetric Gaussian setting in \eqref{eq:cran_3bs_gaussian_channel}, we obtain the following lower bound.
\begin{proposition}\label{proposition:cran_3bs_gaussian_lower_bound}
	The rate $ R $ is achievable if it satisfies the following constraints for some non-negative parameter $ \rho $, $ 0\leq \rho \leq 1 $, $ 0 \leq \rho_c \leq \min \left\{
	\sqrt{\frac{1+\rho}{2}}, \sqrt{\frac{1+2\rho}{3}}
	\right\} $:
	\begin{align} \label{eq:lower_bound_gaussian_symmetric1}
		R < \min\left\{
		\begin{aligned}
			&3C - F_1,
			C + 2C_0 + F_2, \\
			&2C + 4C_0 + F_3 -F_4, \\
			& F_5, 
			 \frac{1}{2} \left[
			C + 2C_0 + F_2+F_6
			\right], \\
			& C + 2C_0 + \frac{1}{2} \left[
			F_3+F_6 - F_4
			\right],  \\
			& C + 2C_0 + \frac{1}{3} \left[
			2F_6 - F_1
			\right] .
		\end{aligned}
		\right\}
	\end{align}
\end{proposition}
The proof of this proposition is relegated to Appendix \ref{appendix:cran_3bs_gaussian_lower_bound_proof}.

%
%
For the upper bound, we choose $ \rv{V} $ to be a noisy version of $ \rv{Y} $, i.e., $ \rv{U} = \rv{Y} + \rv{W}  $ where $ \rv{W} \sim (0,N) $. 
We then specialize \autoref{theorem:cran_3bs_upper_bound2} for the symmetric Gaussian setting here and obtain the following upper bound.
\begin{proposition}\label{proposition:cran_3bs_upper_bound2_gaussian}
	Rate $ R $ is achievable only if it satisfies the following constraints for some $ 0\leq \rho, \rho_c \leq 1 $, such that $ 0 \leq \rho_c \leq \min \left\{
	\sqrt{\frac{1+\rho}{2}}, \sqrt{\frac{1+2\rho}{3}}
	\right\} $
	\begin{equation} \label{eq:cranSym_3bs_upper_bound2_gaussian}
		R < \min\left\{
		\begin{aligned}
			&3C ,
			C + 2C_0 + F_2, \\
			&2C + 4C_0 + F_3,
			 F_5 ,
			3C  + F_7.
		\end{aligned}
		\right\}
	\end{equation}
\end{proposition}
The proof of this proposition is relegated to Appendix \ref{appendix:cran_3bs_gaussian_upper_bound2_proof}.

We provide an evaluation of the proposed bounds from Props. \ref{proposition:cran_3bs_gaussian_lower_bound},
and  \ref{proposition:cran_3bs_upper_bound2_gaussian} for $ P=1 $  in \autoref{figure:cran_3bs_bounds_p_1}.

\section{Conclusions and Discussion}
This paper examines the contribution of cooperation on a diamond network with two and three conferencing relays. Our upper bound presented tighter results than the prior art for the two relays scenario. Furthermore, we proposed the three conferencing base stations scenario. A new coding technique has been developed, and upper bounds were derived, which were shown to outperform the cut-set bound for a regime of channel parameters.

Extending the technique proposed here to an arbitrary $K$ relays setting is challenging due to the higher complexity of multiple paths and loops of information transmission in a cooperative network. However, it is manageable once some symmetry structure is assumed and restrictions on the cooperation links are imposed, and this is the focus of our future work.
%
%

%
	
\appendix

\subsection{Multiletter Upper bound}
We begin our discussion here by presenting a multi-letter expression for the upper bound that will be utilized in the sequence to establish our single-letter upper bounds of \autoref{theorem:cran_2bs_upper_bound1}, \autoref{theorem:cran_2bs_upper_bound2} and \autoref{theorem:cran_3bs_upper_bound2}.
Denote $ \rv{W}_c \triangleq (\rv{W}_{12},\rv{W}_{13},\rv{W}_{21},\rv{W}_{23},\rv{W}_{31},\rv{W}_{32}) $.
\begin{theorem}\label{theorem:multiletter_upper_bound}
    Rate $R$ is achievable if only it satisfies the following list of inequalities
	\begin{align}
		R & \leq \sum_{\omega \in [3]} C_{\omega \omega} -  \frac{1}{n} \Gamma(\rv{X}_{[3]}^n|\rv{W}_{c}),
		\label{eq:multiletter_upper_bound1} \\
		R &\leq \sum_{ \omega \in \set{S}} C_{\omega \omega} + \sum_{\omega \in [3]}  \sum_{\omega' \neq \omega \in \set{S}^c} C_{\omega\omega'} +  \frac{1}{n} \label{eq:multiletter_upper_bound2} I(\rv{X}_{\set{S}^c}^n;\rv{Y}^n|\rv{W}_c,\rv{X}_{\set{S}}^n) ,\\
		R &\leq  \frac{1}{n} I(\rv{X}_{[3]}^n;\rv{Y}^n)
		\label{eq:multiletter_upper_bound5}.
	\end{align}
	for every $ \set{S} \subset [3]  $ and some distribution $\mathsf{P}_{\rv{W}_c \rv{X}_{[3]}^n}$ where $n$ is sufficiently large.
\end{theorem}


\begin{proof}
The following lemma will be useful in deriving a multiletter upper bound.
\begin{lemma} \label{lemma:dpi_cran_conf}
	For the setup defined in \autoref{figure:3bs_1mu_block_diagram}, we have the following type of Data-Processing Inequality.
	\begin{equation}\label{eq:dpi_cran_conf}
		\Gamma(\rv{X}_1^n;\rv{X}_2^n;\rv{X}_3^n|\rv{W}_{c})   \leq \Gamma(\rv{W}_1;\rv{W}_2;\rv{W}_3) .
	\end{equation}
\end{lemma}
The proof of this lemma is deferred to Appendix \ref{appendix:dpi_cran_conf_proof}.

%

By Fano's inequality \cite[Theorem 2.10.1]{Cover2006},
\begin{align}
	nR &= H(\rv{M}) \\
	&= I(\rv{M};\hat{\rv{M}}) + H(\rv{M}|\hat{\rv{M}}) \\
	&\leq I(\rv{M};\rv{Y}^n) + n \epsilon_n,
\end{align}
where $ \epsilon_n $ tends to zero as $ n \rightarrow \infty $.  Further, applying \acrfull{dpi} \cite{Cover2006} on the Markov chain $ \rv{M} \rightarrow (\rv{W}_{[3]}) \rightarrow \rv{Y}^n $, we obtain:
\begin{align}
	I(\rv{M};\rv{Y}^n)
	&\leq I(\rv{W}_{[3]};\rv{Y}^n) \\
	&\eqann{\leq}{a} H(\rv{W}_{[3]}) \\
	&= \sum_{\omega \in [3]} H(\rv{W}_\omega)  - \Gamma(\rv{W}_{[3]}) \\
	&\leq n\sum_{\omega \in [3]} C_{\omega \omega}  - \Gamma(\rv{W}_{[3]}) ,
\end{align}
where \eqannref{a} follows from non-negativity of entropy. 
We proceed to lower bound on $ \Gamma(\rv{W}_{[3]})  $  incorporating \autoref{lemma:dpi_cran_conf}. Using the inequality in \eqref{eq:dpi_cran_conf}, we obtain.
\begin{equation}
	nR \leq n\sum_{\omega \in \Omega} C_{\omega \omega} -  \Gamma(\rv{X}_{[3]}^n|\rv{W}_{c}) + n \epsilon_n.
\end{equation}

In addition, for every $ \set{S} \subseteq [3] $, we have the following list of inequalities:
\begin{align}
	I(\rv{M};\rv{Y}^n)
	&\leq I(\rv{W}_{[3]};\rv{Y}^n) \\
	&= I(\rv{W}_{\set{S}},\rv{W}_{\set{S}^c};\rv{Y}^n) \\
	&= I(\rv{W}_{\set{S}},\rv{W}_{\set{S}^c},\rv{W}_c;\rv{Y}^n) \\
	&= I(\rv{W}_{\set{S}};\rv{Y}^n)
	+  I(\rv{W}_c;\rv{Y}^n|\rv{W}_{\set{S}})
	+  I(\rv{W}_{\set{S}^c};\rv{Y}^n|\rv{W}_{\set{S}},\rv{W}_c) \\
	&\leq H(\rv{W}_{\set{S}}) + H(\rv{W}_c|\rv{W}_{\set{S}})  + 
	I(\rv{X}_{\set{S}^c}^n;\rv{Y}^n|\rv{X}_{\rv{W}_c,\set{S}}^n) \\
	& \leq n \sum_{ \omega \in \set{S}} C_{\omega \omega} + \sum_{\omega \in [3]}  \sum_{\omega' \neq \omega \in \set{S}^c} C_{\omega \omega'} + I(\rv{X}_{\set{S}^c}^n;\rv{Y}^n|\rv{W}_c,\rv{X}_{\set{S}}^n).
\end{align}
Thus,
\begin{equation}
	nR \leq n \sum_{ \omega \in \set{S}} C_{\omega \omega } + \sum_{\omega \in [3]}  \sum_{\omega' \neq \omega \in \set{S}^c} C_{\omega \omega'} + I(\rv{X}_{\set{S}^c}^n;\rv{Y}^n|\rv{W}_c,\rv{X}_{\set{S}}^n) + n \epsilon_n.
\end{equation}

Moreover, also using \acrshort{dpi} for the Markov chain $\rv{M} \rightarrow \rv{X}_{[3]}^n \rightarrow \rv{Y}^n$ , we obtain
\begin{equation}
	nR \leq I(\rv{M} ;\rv{Y}^n) + n \epsilon_n \leq I(\rv{X}_{[3]}^n ;\rv{Y}^n) + n \epsilon_n.
\end{equation}
This completes the proof of the multiletter upper bound.
\end{proof}

	\subsection{Auxiliary Lemma} \label{appendix:auxiliary_lemma_proof}
	\begin{lemma} \label{lemma:auxiliary_lemma}
	$ I(\rv{X};\rv{Y}|\rv{Z}) $ can be written in the following form for any random variable $ \rv{U} $:
	\begin{equation}
		I(\rv{X};\rv{Y}|\rv{Z}) 
		= I(\rv{X},\rv{Y};\rv{U}|\rv{Z}) -I(\rv{Y};\rv{U}|\rv{X},\rv{Z})  +I(\rv{X};\rv{Y}|\rv{U},\rv{Z})  - I(\rv{X};\rv{U}|\rv{Y},\rv{Z}) .
	\end{equation}
\end{lemma}
\begin{proof}
	\begin{align}
	I(\rv{X};\rv{Y}|\rv{Z}) 
	& = I(\rv{X};\rv{Y},\rv{U}|\rv{Z})  - I(\rv{X};\rv{U}|\rv{Y},\rv{Z}) \\
	& = I(\rv{X};\rv{U}|\rv{Z}) + I(\rv{X};\rv{Y}|\rv{U},\rv{Z})  - I(\rv{X};\rv{U}|\rv{Y},\rv{Z}) \\
	& = I(\rv{X},\rv{Y};\rv{U}|\rv{Z}) -I(\rv{Y};\rv{U}|\rv{X},\rv{Z})  +I(\rv{X};\rv{Y}|\rv{U},\rv{Z})  - I(\rv{X};\rv{U}|\rv{Y},\rv{Z}) .
	\end{align}
\end{proof}
	
	\subsection{Proof of \autoref{theorem:cran_2bs_upper_bound1}} \label{appendix:cran_2bs_upper_bound1_proof}
	Consider the multi-letter upper bound from \autoref{theorem:multiletter_upper_bound} with $\set{X}_3 = \set{W}_{13}= \set{W}_{23} = \set{W}_{33} = \set{W}_{31} = \set{W}_{32} = \emptyset$, $C_{33} = C_{23}=C_{13} = C_{32} = C_{31} = 0$ and $C_{11} = C_1 $. $C_2 = C_{22}$ . Plugging $ \rv{X} \mapsto \rv{X}_1^n $, $ \rv{Y} \mapsto \rv{X}_2^n $, $ \rv{Z} \mapsto (\rv{W}_{12},\rv{W}_{21}) $  and $ \rv{U} \mapsto \rv{V}^n $ in \autoref{lemma:auxiliary_lemma}, we obtain:
\begin{align}
	I(\rv{X}_1^n;\rv{X}_2^n|\rv{W}_{12},\rv{W}_{21}) 
	&= I(\rv{X}_1^n,\rv{X}_2^n;\rv{V}^n|\rv{W}_{12},\rv{W}_{21}) - I(\rv{X}_{2};\rv{V}^n|\rv{W}_{12},\rv{W}_{21},\rv{X_1^n})  \\
	&+ I(\rv{X}_1^n;\rv{X}_2^n|\rv{W}_{12},\rv{W}_{21},\rv{V}^n) - I(\rv{X}_1^n;\rv{V}^n|\rv{W}_{12},\rv{W}_{21},\rv{X}_2^n)\\
	&\geq I(\rv{X}_1^n,\rv{X}_2^n;\rv{V}^n|\rv{W}_{12},\rv{W}_{21}) - I(\rv{X}_{2};\rv{V}^n|\rv{W}_{12},\rv{W}_{21},\rv{X_1^n})  
	 - I(\rv{X}_1^n;\rv{V}^n|\rv{W}_{12},\rv{W}_{21},\rv{X}_2^n),
\end{align}
where $ \rv{V}^n $ is an arbitrary random sequence.
Thus \eqref{eq:multiletter_upper_bound1} in \autoref{theorem:multiletter_upper_bound} can be expanded as follows:
\begin{align} 
	nR & \leq n(C_1+C_2)-  I(\rv{X}_1^n;\rv{X}_2^n|\rv{W}_{12},\rv{W}_{21}) \\
	&\leq n(C_1+C_2) -  I(\rv{X}_1^n,\rv{X}_2^n;\rv{V}^n|\rv{W}_{12},\rv{W}_{21}) 
	+ I(\rv{X}_{2}^n;\rv{V}^n|\rv{W}_{12},\rv{W}_{21},\rv{X_1^n})  
	+ I(\rv{X}_1^n;\rv{V}^n|\rv{W}_{12},\rv{W}_{21},\rv{X}_2^n).
	\label{eq:upper_bound1_ratebound1}
\end{align}
The most challenging term here to bound from above is the minus of mutual information. We will eliminate it, in a similar manner as was suggested in \cite[Proof of Theorem 2]{SaeediBidokhti2016}. Note that the multi-letter bound given in \eqref{eq:multiletter_upper_bound2}, with $\set{S} = \emptyset$, can be further bounded from above as
\begin{equation} \label{eq:upper_bound1_rateboundX}
	nR \leq n(C_{12} + C_{21}) + I(\rv{X}_1^n, \rv{X}_2^n;\rv{Y}^n|\rv{W}_{21},\rv{W}_{12})
	\leq n(C_{12} + C_{21}) + I(\rv{X}_1^n, \rv{X}_2^n;\rv{Y}^n,\rv{V}^n|\rv{W}_{21},\rv{W}_{12}). 
\end{equation}
Combining the inequalities \eqref{eq:upper_bound1_ratebound1} and \eqref{eq:upper_bound1_rateboundX} we obtain
\begin{equation}\label{eq:cran_upper_bound_combination}
	2nR \leq n(C_1+C_2+C_{21} +  C_{12}) + I(\rv{X}_1^n, \rv{X}_2^n;\rv{Y}^n|\rv{W}_{21},\rv{W}_{12},\rv{V}^n) +  I(\rv{X}_{2}^n;\rv{V}^n|\rv{W}_{12},\rv{W}_{21},\rv{X_1^n})  
	+ I(\rv{X}_1^n;\rv{V}^n|\rv{W}_{12},\rv{W}_{21},\rv{X}_2^n).
\end{equation}
Furthermore, since $ \rv{V}^n $ is arbitrary, we define $ 
\rv{V}_i $ from $ \rv{X}_{1i},\rv{X}_{2i}, \rv{Y}_i $ through the channel $ \mathsf{P}_{\rv{V}|\rv{X}_1,\rv{X}_2,\rv{Y}}(v_i|x_{1i},x_{2i},y_i) $, $ i=1,2,\dots,n $, in similar manner as was done in \cite{SaeediBidokhti2016}. 
We can expand the first mutual information term in \eqref{eq:cran_upper_bound_combination} as follows:
\begin{align}
	I(\rv{X}_1^n,\rv{X}_2^n;\rv{Y}^n|\rv{W}_{12},\rv{W}_{21},\rv{V}^n) 
	&= \sum_{i=1}^n I(\rv{X}_1^n,\rv{X}_2^n;\rv{Y}_i|\rv{W}_{12},\rv{W}_{21},\rv{V}^n,\rv{Y}^{i-1}) \\
	&= \sum_{i=1}^n I(\rv{X}_{1i},\rv{X}_{2i};\rv{Y}_i|\rv{W}_{12},\rv{W}_{21},\rv{V}^n,\rv{Y}^{i-1}) \\
	&\leq \sum_{i=1}^n I(\rv{X}_{1i},\rv{X}_{2i};\rv{Y}_i|\rv{W}_{12},\rv{W}_{21},\rv{V}_i)
\end{align}
where the last equality follows due to the following Markov chain:
\begin{equation}
	(\rv{X}_{1}^n,\rv{X}_2^n,\rv{Y}^{i-1},\rv{V}^n) 
	\rightarrow 
	(\rv{X}_{1i},\rv{X}_{2i},\rv{V}_i) \rightarrow \rv{Y}_i.
\end{equation}
As was suggested in \cite{Wigger2008},  We introduce $ \tilde{\rv{U}}_i \triangleq (\rv{W}_{12},\rv{W}_{21}) $.  
With this choice of $ \tilde{\rv{U}}_i $ and $ \rv{V}_i $, and utilizing the following Markov chains
\begin{align}
	(\rv{X}_{1}^n,\rv{X}_2^n,\tilde{\rv{U}}_i,\rv{Y}^{i-1},\rv{V}^n) &\rightarrow (\tilde{\rv{U}}_i,\rv{X}_{1i},\rv{X}_{2i},\rv{V}_i) \rightarrow \rv{Y}_i, \\
	(\rv{X}_{1}^n,\rv{X}_2^n,\tilde{\rv{U}}_i,\rv{V}^{i-1}) &\rightarrow (\tilde{\rv{U}}_i,\rv{X}_{1i},\rv{X}_{2i}) \rightarrow \rv{V}_i.
\end{align}
Those Markov chains follow from the following considerations:
\begin{align}
	\mathsf{P}(v^n|w_{12},w_{21},x_1^n,x_2^n) 
	&= \prod_{i=1}^n  \mathsf{P}(v_i|w_{12},w_{21},x_1^n,x_2^n,v^{i-1})  \\
	&= \prod_{i=1}^n \sum_{y_i \in \set{Y}} \mathsf{P}(v_i,y_i|w_{12},w_{21},x_1^n,x_2^n,v^{i-1})  \\
	&= \prod_{i=1}^n \sum_{y_i \in \set{Y}} \mathsf{P}(v_i|w_{12},w_{21},x_1^n,x_2^n,y_i,v^{i-1}) \mathsf{P}(y_i|w_{12},w_{21},x_1^n,x_2^n,v^{i-1}) \\
	&= \prod_{i=1}^n \sum_{y_i \in \set{Y}} \mathsf{P}(v_i|u_i,x_{1i},x_{2i},y_i) \mathsf{P}(y_i|x_{1i},x_{2i}) \\
	&= \prod_{i=1}^n \sum_{y_i \in \set{Y}} \mathsf{P}(v_i|u_i,x_{1i},x_{2i},y_i) \mathsf{P}(y_i|u_i,x_{1i},x_{2i}) \\
	&= \prod_{i=1}^n \sum_{y_i \in \set{Y}} \mathsf{P}(v_i|u_i,x_{1i},x_{2i}),
\end{align}
and
\begin{align}
	\mathsf{P}(y^n,v^n|x_1^n,x_2^n,w_{12},w_{21}) 
	&= \prod_{i=1}^n  \mathsf{P}(v_i|w_{12},w_{21},x_{1i},x_{2i},y_i) \mathsf{P}(y_i|x_{1i},x_{2i}) \\
	&= \prod_{i=1}^n  \mathsf{P}(v_i|w_{12},w_{21},x_{1i},x_{2i},y_i) \mathsf{P}(y_i|w_{12},w_{21},x_{1i},x_{2i}) \\
	&= \prod_{i=1}^n  \mathsf{P}(y_i,v_i|w_{12},w_{21},x_{1i},x_{2i}) \\
	&= \prod_{i=1}^n  \mathsf{P}(v_i|w_{12},w_{21},x_{1i},x_{2i}) \mathsf{P}(y_i|w_{12},w_{21},x_{1i},x_{2i},v_i) \\
	&= \mathsf{P}(v^n|x_1^n,x_2^n,w_{12},w_{21})  \mathsf{P}(y^n|x_1^n,x_2^n,w_{12},w_{21},v^n) \\
	&= \prod_{i=1}^n  \mathsf{P}(v_i|w_{12},w_{21},x_{1i},x_{2i}) \mathsf{P}(y_i|w_{12},w_{21},x_{1}^n,x_{2}^n,v^n,y^{i-1}).
\end{align}
Thus, we obtain
\begin{align}
	I(\rv{X}_1^n,\rv{X}_2^n;\rv{Y}^n|\rv{W}_{12},\rv{W}_{21},\rv{V}^n) 
	&\leq \sum_{i=1}^n I(\rv{X}_{1i},\rv{X}_{2i};\rv{Y}_i|\tilde{\rv{U}}_i,\rv{V}_i) \\
	&= n I(\rv{X}_{1\rv{Q}},\rv{X}_{2 \rv{Q}};\rv{Y}_{\rv{Q}}|\tilde{\rv{U}}_{\rv{Q}},\rv{V}_{\rv{Q}},\rv{Q}).
\end{align}
Similarly,
\begin{align}
	I(\rv{X}_{2}^n;\rv{V}^n|\rv{W}_{12},\rv{W}_{21},\rv{X}_1^n)  
	&= \sum_{i=1}^n I(\rv{X}_{2}^n;\rv{V}_i|\rv{W}_{12},\rv{W}_{21},\rv{X}_1^n,\rv{V}^{i-1})  \\
	&\leq \sum_{i=1}^n I(\rv{X}_{2i};\rv{V}_i|\tilde{\rv{U}}_i,\rv{X_{1i}})  \\
	&= n I(\rv{X}_{2\rv{Q}};\rv{V}_{\rv{Q}}|\tilde{\rv{U}}_{\rv{Q}},\rv{X_{1\rv{Q}}},\rv{Q}),
\end{align}
and,
\begin{equation}
	I(\rv{X}_1^n;\rv{V}^n|\rv{W}_{12},\rv{W}_{21},\rv{X}_2^n)
	\leq
	n I(\rv{X}_{1\rv{Q}};\rv{V}_{\rv{Q}}|\tilde{\rv{U}}_{\rv{Q}},\rv{X_{2\rv{Q}}},\rv{Q}).
\end{equation}
Denoting $ \tilde{\rv{U}} = \tilde{\rv{U}}_{\rv{Q}} $, $ \rv{X}_1 = \rv{X}_{1\rv{Q}} $, $ \rv{X}_2 = \rv{X}_{2\rv{Q}} $, and $ \rv{V} = \rv{V}_{\rv{Q}} $,  we obtain the following single-letter upper bound on $R$,
\begin{equation}
	2R \leq  C_1 + C_2 +  C_{21}+C_{12} + I(\rv{X}_{1},\rv{X}_{2 };\rv{Y}|\tilde{\rv{U}},\rv{V},\rv{Q}) + I(\rv{X}_1;\rv{V}|\tilde{\rv{U}},\rv{X}_2,\rv{Q}) + I(\rv{X}_2;\rv{V}|\tilde{\rv{U}},\rv{X}_1,\rv{Q}).
\end{equation}

As for \eqref{eq:multiletter_upper_bound2} with $\set{S} = \{1\}$, we have
\begin{align*}
	I(\rv{X}_2^n;\rv{Y}^n|\rv{W}_{12},\rv{W}_{21},\rv{X}_1^n)
	&= \sum_{i=1}^n I(\rv{X}_2^n;\rv{Y}_i|\rv{W}_{12},\rv{W}_{21},\rv{X}_1^n,\rv{Y}^{i-1}) \\
	&\leq \sum_{i=1}^n I(\rv{X}_{2i};\rv{Y}_i|\tilde{\rv{U}}_{i},\rv{X}_{1i}) \\
	&=  n I(\rv{X}_{2\rv{Q}};\rv{Y}_{\rv{Q}}| \tilde{\rv{U}}_{\rv{Q}}, \rv{X}_{1\rv{Q}},\rv{Q}) \\
	&=  n I(\rv{X}_2;\rv{Y}|\tilde{\rv{U}},\rv{X}_{1},\rv{Q}).
\end{align*}
In a very similar manner, one can show that
\begin{equation}
	I(\rv{X}_1^n;\rv{Y}^n|\rv{W}_{12},\rv{W}_{21},\rv{X}_2^n) \leq n I(\rv{X}_1;\rv{Y}|\tilde{\rv{U}},\rv{X}_{2},\rv{Q}),
\end{equation}
and
\begin{equation}
	I(\rv{X}_1^n,\rv{X}_2^n;\rv{Y}^n|\rv{W}_{12},\rv{W}_{21}) \leq n I(\rv{X}_1,\rv{X}_{2};\rv{Y}|\tilde{\rv{U}},\rv{Q}).
\end{equation}
and
\begin{equation}
	I(\rv{X}_1^n,\rv{X}_2^n;\rv{Y}^n) \leq n I(\rv{X}_1,\rv{X}_{2};\rv{Y}|\rv{Q}).
\end{equation}
Finally, note that $I(\rv{X}_1,\rv{X}_{2};\rv{Y}|\rv{Q}) \leq I(\rv{X}_1,\rv{X}_{2},\rv{Q};\rv{Y}) = I(\rv{X}_1,\rv{X}_{2};\rv{Y})$, thus we may define $\rv{U} \triangleq (\tilde{\rv{U}},\rv{Q})$ and 
this concludes the proof of the theorem.

	\subsection{Proof of \autoref{remark:upper_bound_equivalent_form}}
	\label{appendix:remark_upper_bound_proof}
	Using the chain rule property of mutual information we have
\begin{align}
	&I(\rv{X}_{1},\rv{X}_{2};\rv{Y}|\rv{U},\rv{V})+ I(\rv{X}_1;\rv{V}|\rv{U},\rv{X}_2) +I(\rv{X}_2;\rv{V}|\rv{U},\rv{X}_1) \\
	&= I(\rv{X}_{1},\rv{X}_{2};\rv{Y},\rv{V}|\rv{U}) - I(\rv{X}_{1},\rv{X}_{2};\rv{V}|\rv{U})
	+ I(\rv{X}_1,\rv{X}_2;\rv{V}|\rv{U}) \\
	&- I(\rv{X}_2;\rv{V}|\rv{U})
	 + I(\rv{X}_2;\rv{X}_1,\rv{V}|\rv{U})
	 - I(\rv{X}_2;\rv{X_1}|\rv{U}) \\
	 &= I(\rv{X}_{1},\rv{X}_{2};\rv{Y},\rv{V}|\rv{U}) 
	 + I(\rv{X}_1;\rv{X}_2|\rv{U},\rv{V})
	 - I(\rv{X}_2;\rv{X_1}|\rv{U}).
\end{align}
Also, since we choose $ \mathsf{P}_{\rv{V}|\rv{U} \rv{X}_1 \rv{X}_2 \rv{Y}} = \mathsf{P}_{\rv{V}|\rv{Y}} $, it implies the following Markov chain:
\begin{equation}
	\rv{U} \rightarrow (\rv{X}_1,\rv{X}_2) \rightarrow \rv{Y} \rightarrow \rv{V},
\end{equation}
and thus
\begin{equation}
	I(\rv{X}_{1},\rv{X}_{2};\rv{Y},\rv{V}|\rv{U}) = I(\rv{X}_{1},\rv{X}_{2};\rv{Y}|\rv{U}) .
\end{equation}
	
	\subsection{Proof of \autoref{theorem:cran_2bs_upper_bound2}} \label{appendix:cran_2bs_upper_bound2_proof}
	From Appendix \ref{appendix:cran_2bs_upper_bound1_proof}, \eqref{eq:multiletter_upper_bound1} in \autoref{theorem:multiletter_upper_bound} can be expanded as follows:
\begin{equation}
	nR \leq n(C_1+C_2) -  I(\rv{X}_1^n,\rv{X}_2^n;\rv{V}^n|\rv{W}_{12},\rv{W}_{21}) 
	+ I(\rv{X}_{2}^n;\rv{V}^n|\rv{W}_{12},\rv{W}_{21},\rv{X_1^n})  
	+ I(\rv{X}_1^n;\rv{V}^n|\rv{W}_{12},\rv{W}_{21},\rv{X}_2^n).
\end{equation}
As was suggested in \cite{SaeediBidokhti2016}, since $ \rv{V}^n $ is arbitrary, we choose each symbol $ 
\rv{V}_i $ as the output of the channel  $ \mathsf{P}_{\rv{V}|\rv{Y}}(v_i|y_i) $, with input $ y_i $, $ i=1,2,\dots,n $. We have,
\begin{align}
	\mathsf{P}_{\rv{V}^n|\rv{X}_1^n,\rv{X}_2^n} (v^n|x_1^n,x_2^n)
	&= \sum_{y^n \in \set{Y}^n} \mathsf{P}_{\rv{V}^n|\rv{Y}^n} (v^n|y^n) \mathsf{P}_{\rv{Y}^n|\rv{X}_1^n,\rv{X}_2^n} (v^n|x_1^n,x_2^n) \\
	&= \sum_{y^n \in \set{Y}^n} \prod_{i=1}^n \mathsf{P}_{\rv{V}|\rv{Y}}(v_i|y_i)  \mathsf{P}_{\rv{Y}|\rv{X}_1\rv{X}_2}(y_i|x_{1i},x_{2i}) \\
	&= \prod_{i=1}^n \mathsf{P}_{\rv{V}|\rv{X}_1\rv{X}_2}(v_i|x_{1i},x_{2i}).
\end{align}

Furthermore, we denote $ \tilde{\rv{U}}_i = (\rv{W}_{12},\rv{W}_{21}) $. With this choice of $ \rv{V}_i $ and $ \tilde{\rv{U}}_i $, we can expand $ I(\rv{X}_1^n,\rv{X}_2^n;\rv{V}^n|\rv{W}_{12},\rv{W}_{21})  $ as follows:
\begin{align}
	I(\rv{X}_1^n,\rv{X}_2^n;\rv{V}^n|\rv{W}_{12},\rv{W}_{21}) 
	&= \sum_{i=1}^n I(\rv{X}_1^n,\rv{X}_2^n;\rv{V}_i|\rv{W}_{12},\rv{W}_{21},\rv{V}^{i-1}) \\
	&= \sum_{i=1}^n I(\rv{X}_{1i},\rv{X}_{2i};\rv{V}_i|\tilde{\rv{U}}_{i},\rv{V}^{i-1}) \\
	&= n I(\rv{X}_{1\rv{Q}},\rv{X}_{2 \rv{Q}};\rv{V}_{\rv{Q}}|\tilde{\rv{U}}_{\rv{Q}},\rv{V}^{\rv{Q}-1},\rv{Q}).
\end{align}
Also,
\begin{align}
	I(\rv{X}_{2}^n;\rv{V}^n|\rv{W}_{12},\rv{W}_{21},\rv{X_1^n})  
	&= \sum_{i=1}^n I(\rv{X}_{2}^n;\rv{V}_i|\rv{W}_{12},\rv{W}_{21},\rv{X_1^n},\rv{V}^{i-1})  \\
	&\leq \sum_{i=1}^n I(\rv{X}_{2i};\rv{V}_i|\tilde{\rv{U}}_i,\rv{X_{1i}},\rv{V}^{i-1})  \\
	&\leq n I(\rv{X}_{2\rv{Q}};\rv{V}_{\rv{Q}}|\tilde{\rv{U}}_{\rv{Q}},\rv{X_{1\rv{Q}}},\rv{V}^{\rv{Q}-1},\rv{Q}).
\end{align}
Similarly,
\begin{equation}
	I(\rv{X}_1^n;\rv{V}^n|\rv{W}_{12},\rv{W}_{21},\rv{X}_2^n)
	\leq
	n I(\rv{X}_{1\rv{Q}};\rv{V}_{\rv{Q}}|\tilde{\rv{U}}_{\rv{Q}},\rv{X_{2\rv{Q}}},\rv{V}^{\rv{Q}-1},\rv{Q}).
\end{equation}
Combining the above inequalities we obtain:
\begin{equation}
	R \leq C_1+C_2 - I(\rv{X}_{1\rv{Q}},\rv{X}_{2\rv{Q}};\rv{V}_{\rv{Q}}|\tilde{\rv{U}}_{\rv{Q}},\rv{V}^{\rv{Q}-1},\rv{Q}) + I(\rv{X}_{1\rv{Q}};\rv{V}_{\rv{Q}}|\tilde{\rv{U}}_{\rv{Q}},\rv{V}^{\rv{Q}-1},\rv{X}_{2\rv{Q}},\rv{Q}) 
	+ I(\rv{X}_{2\rv{Q}};\rv{V}_{\rv{Q}}|\tilde{\rv{U}}_{\rv{Q}},\rv{V}^{\rv{Q}-1},\rv{X}_{1\rv{Q}},\rv{Q}) .
\end{equation}

Next, consider the multi-letter mutual information term in the \acrshort{rhs} of \eqref{eq:multiletter_upper_bound2} with $\set{S} = \{1\}$,
\begin{align}
	 I(\rv{X}_2^n;\rv{Y}^n|\rv{W}_{12},\rv{W}_{21},\rv{X}_1^n)
	 &= \sum_{i=1}^n  I(\rv{X}_2^n;\rv{Y}_i|\rv{W}_{12},\rv{W}_{21},\rv{X}_1^n,\rv{Y}^{i-1}) \\
	 &\leq \sum_{i=1}^n  I(\rv{X}_{2i};\rv{Y}_i|\tilde{\rv{U}}_{i},\rv{X}_{1i},\rv{Y}^{i-1}) \\
	 &\eqann{=}{a} \sum_{i=1}^n  I(\rv{X}_{2i};\rv{Y}_i|\tilde{\rv{U}}_{i},\rv{X}_{1i},\rv{Y}^{i-1},\rv{V}^{i-1}) \\
	 & \leq \sum_{i=1}^n  I(\rv{X}_{2i};\rv{Y}_i|\tilde{\rv{U}}_{i},\rv{X}_{1i},\rv{V}^{i-1}) \\
	 & = n  I(\rv{X}_{2 \rv{Q}};\rv{Y}_{\rv{Q}}|\tilde{\rv{U}}_{\rv{Q}},\rv{X}_{1\rv{Q}},\rv{V}^{\rv{Q}-1},\rv{Q}) ,
\end{align}
where \eqannref{a} follows since $ \rv{V}^n $ is the output of $ \prod_{i=1}^n \mathsf{P}_{\rv{V}|\rv{Y}}(v_i|y_i)  $, therefore conditioning on $\rv{V}^{i-1}$ does not adds information given $\rv{Y}^{i-1}$.
Similarly, one can show that
\begin{equation}
	 I(\rv{X}_1^n;\rv{Y}^n|\rv{W}_{12},\rv{W}_{21},\rv{X}_2^n)
	 \leq n  I(\rv{X}_{1 \rv{Q}};\rv{Y}_{\rv{Q}}|\tilde{\rv{U}}_{\rv{Q}},\rv{X}_{2\rv{Q}},\rv{V}^{\rv{Q}-1},\rv{Q}),
\end{equation}
and
\begin{equation}
	I(\rv{X}_1^n,\rv{X}_2^n;\rv{Y}^n|\rv{W}_{12},\rv{W}_{21})
	\leq n  I(\rv{X}_{1 \rv{Q}},\rv{X}_{2\rv{Q}};\rv{Y}_{\rv{Q}}|\tilde{\rv{U}}_{\rv{Q}},\rv{V}^{\rv{Q}-1},\rv{Q}).
\end{equation}

To this end we obtained the following list of inequalities:
\begin{align}
	R & \leq C_1+C_2 - I(\rv{X}_{1\rv{Q}},\rv{X}_{2\rv{Q}};\rv{V}_{\rv{Q}}|\tilde{\rv{U}}_{\rv{Q}},\rv{V}^{\rv{Q}-1},\rv{Q}) + I(\rv{X}_{1\rv{Q}};\rv{V}_{\rv{Q}}|\tilde{\rv{U}}_{\rv{Q}},\rv{V}^{\rv{Q}-1},\rv{X}_{2\rv{Q}},\rv{Q}), \\
	&+ I(\rv{X}_{2\rv{Q}};\rv{V}_{\rv{Q}}|\tilde{\rv{U}}_{\rv{Q}},\rv{V}^{\rv{Q}-1},\rv{X}_{1\rv{Q}},\rv{Q}) , \\
	R & \leq C_1+C_{12} +  I(\rv{X}_{2\rv{Q}};\rv{Y}_{\rv{Q}}|\tilde{\rv{U}}_{\rv{Q}},\rv{X}_{1\rv{Q}},\rv{V}^{\rv{Q}-1},\rv{Q}), \\
	R & \leq C_{2} + C_{21} +  I(\rv{X}_{1\rv{Q}};\rv{Y}_{\rv{Q}}|\tilde{\rv{U}}_{\rv{Q}},\rv{X}_{2 \rv{Q}},\rv{V}^{\rv{Q}-1},\rv{Q}), \\
	R & \leq I(\rv{X}_{1\rv{Q}},\rv{X}_{2\rv{Q}};\rv{Y}_{\rv{Q}}|\rv{Q}), \\
	R &\leq C_{12} + C_{21} +  I(\rv{X}_{1\rv{Q}},\rv{X}_{2\rv{Q}};\rv{Y}_{\rv{Q}}|\tilde{\rv{U}}_{\rv{Q}},\rv{V}^{\rv{Q}-1}, \rv{Q}).
\end{align}
The theorem follows after observing that $ I(\rv{X}_{1 \rv{Q}}, \rv{X}_{2\rv{Q}};\rv{Y}_{\rv{Q}}|\rv{Q}) \leq I(\rv{X}_{1 \rv{Q}},\rv{X}_{2\rv{Q}}, \rv{Q};\rv{Y}_{\rv{Q}}) = I(\rv{X}_{1 \rv{Q}},\rv{X}_{2\rv{Q}};\rv{Y}_{\rv{Q}}) $, and denoting $ \rv{X}_1 = \rv{X}_{1\rv{Q}} $, $ \rv{X}_2 = \rv{X}_{2\rv{Q}} $, $ \rv{Y} = \rv{Y}_{\rv{Q}} $, $ \rv{V} = \rv{V}_{\rv{Q}} $,   $ \rv{U} = (\tilde{\rv{U}}_{\rv{Q}},\rv{Q}) $ and $ \rv{T} = \rv{V}^{\rv{Q}-1} $.

	\subsection{Proof of \autoref{proposition:cran_2bs_gaussian_lower_bound}}
	\label{appendix:cran_2bs_gaussian_lower_bound_proof}
	Let $ \Sigma_{\rv{X}_1 \rv{U}  \rv{X}_2} $ a general covariance matrix for the triple $ (\rv{X}_1,\rv{U},\rv{X}_2) $ of the form
\begin{equation} \label{eq:general_covariance_matrix}
	\Sigma_{\rv{X}_1 \rv{U}  \rv{X}_2} =
	\begin{pmatrix}
		P_1 & \rho_1 \sqrt{P_1 P_u} & \rho \sqrt{P_1 P_2} \\
		\rho_1 \sqrt{P_1 P_u} & P_u & \rho_2 \sqrt{P_2 P_u} \\
		\rho \sqrt{P_1 P_2}  & \rho_2 \sqrt{P_2 P_u} & P_2
	\end{pmatrix}.
\end{equation}
For the symmetric case we may restrict the covariance matrix over which optimization takes place, i.e, \eqref{eq:general_covariance_matrix}, to $ \rho_1 = \rho_2 = \rho_c $. 
With this choice, we evaluate \autoref{lemma:cran_2bs_dmc_lower_bound} and obtain the desired bound in \autoref{proposition:cran_2bs_gaussian_lower_bound}.
	
	\subsection{Upper Bound Proof for the 2 relay Gaussian channel}
	\label{appendix:cran_2bs_gaussian_upper_bounds_proof}
	We first recall the following lemma from \cite{Thomas1987}.
\begin{lemma}[{\cite[Lemma 1]{Thomas1987}}] \label{lemma:thomas_lemma1}
	Let $ \rv{X}_1,\rv{X}_2,\dots, \rv{X}_k $ be an arbitrary set of zero-mean random variables with covariance matrix $ K $. Let $ \set{S} $ be any subset of $ \{1,2,\dots,k \} $ and $ \set{S}^c $ be its complement. Then
	\begin{equation}
		h(\rv{X}_{\set{S}}|\rv{X}_{\set{S}^c}) \leq h(\rv{X}_{\set{S}}^{\mathcal{G}}|\rv{X}_{\set{S}^c}^{\mathcal{G}})
	\end{equation}
	where $ (\rv{X}_1^{\mathcal{G}},\rv{X}_2^{\mathcal{G}},\dots, \rv{X}_k^{\mathcal{G}}) \sim \mathcal{N}(0,K) $.
\end{lemma}
Utilizing \autoref{lemma:thomas_lemma1} we obtain the following lemma.
\begin{lemma}\label{lemma:upper_bound_gaussian_function}
	Assume that we choose $ \rv{V} = \rv{Y} + \rv{W} $, where $ \rv{W} \sim \mathcal{N}(0,N) $ and independent of all other random variables in the problem. Then, for every random triple $ (\rv{U},\rv{X}_1,\rv{X}_2) $ with covariance matrix $ \Sigma_{\rv{U}\rv{X}_1\rv{X}_2} $, the mutual information terms in the RHS of \eqref{eq:upper_bound_2relays_noQ} are maximized with the respective Gaussian random variables with covariance matrix  $ \Sigma_{\rv{U}\rv{X}_1\rv{X}_2} $.
\end{lemma}
\begin{proof}
	We will show that every term in the RHS of \eqref{eq:upper_bound_2relays_noQ} is bounded from above by the respective jointly Gaussian choice of random variables. 
	\begin{align}
		I(\rv{X}_2;\rv{Y}|\rv{U},\rv{X}_1)
		&= h(\rv{Y}|\rv{U},\rv{X}_1)-h(\rv{Y}|\rv{U},\rv{X}_1,\rv{X}_2) \\
		&= h(\rv{X}_2+\rv{Z}|\rv{U},\rv{X}_1)-h(\rv{Z}) \\
		&\eqann{\leq}{a} h(\rv{X}_2^{\mathcal{G}}+\rv{Z}|\rv{U}^{\mathcal{G}},\rv{X}_1^{\mathcal{G}})-h(\rv{Z}) \\
		&= I(\rv{X}_2^{\mathcal{G}};\rv{Y}^{\mathcal{G}}|\rv{U}^{\mathcal{G}},\rv{X}_1^{\mathcal{G}}),
	\end{align}
	where \eqannref{a} follows from \autoref{lemma:thomas_lemma1}. Similarly, we have also the following inequalities,
	\begin{align}
		I(\rv{X}_1;\rv{Y}|\rv{U},\rv{X}_2) &\leq I(\rv{X}_1^{\mathcal{G}};\rv{Y}^{\mathcal{G}}|\rv{U}^{\mathcal{G}},\rv{X}_2^{\mathcal{G}}), \\
		I(\rv{X}_1,\rv{X}_2;\rv{Y}|\rv{U}) &\leq I(\rv{X}_1^{\mathcal{G}},\rv{X}_2^{\mathcal{G}};\rv{Y}^{\mathcal{G}}|\rv{U}^{\mathcal{G}}), \\
		I(\rv{X}_1,\rv{X}_2;\rv{Y}|\rv{U},\rv{V}) &\leq I(\rv{X}_1^{\mathcal{G}},\rv{X}_2^{\mathcal{G}};\rv{Y}^{\mathcal{G}}|\rv{U}^{\mathcal{G}},\rv{V}^{\mathcal{G}}), \\
		I(\rv{X}_1,\rv{X}_2;\rv{Y}) &\leq I(\rv{X}_1^{\mathcal{G}},\rv{X}_2^{\mathcal{G}};\rv{Y}^{\mathcal{G}}).
	\end{align}
	Furthermore, since $ \rv{V} = \rv{Y} +  \rv{W}$, we obtain
	\begin{align}
		I(\rv{X}_1;\rv{V}|\rv{U},\rv{X}_2) 
		&= h(\rv{V}|\rv{U},\rv{X}_2)  - h(\rv{V}|\rv{U},\rv{X}_1,\rv{X}_2) \\
		&= h(\rv{V}|\rv{U},\rv{X}_2)  - h(\rv{Y} + \rv{W}|\rv{U},\rv{X}_1,\rv{X}_2) \\
		&= h(\rv{V}|\rv{U},\rv{X}_2)  - h(\rv{X}_1+\rv{X}_2+\rv{Z} + \rv{W}|\rv{U},\rv{X}_1,\rv{X}_2) \\
		&= h(\rv{V}|\rv{U},\rv{X}_2)  - h(\rv{Z} + \rv{W}|\rv{U},\rv{X}_1,\rv{X}_2) \\
		&\eqann{=}{a} h(\rv{V}|\rv{U},\rv{X}_2) - h(\rv{Z} + \rv{W}) \\
		&\eqann{\leq}{b} h(\rv{V}^{\mathcal{G}} |\rv{U}^{\mathcal{G}},\rv{X}_2^{\mathcal{G}})  - h(\rv{Z} + \rv{W}) \\
		&= I(\rv{X}_1^{\mathcal{G}};\rv{V}^{\mathcal{G}}|\rv{U}^{\mathcal{G}},\rv{X}_2^{\mathcal{G}}), 
	\end{align}
	where \eqannref{a} follows by requiring that $ \rv{W} $ is independent of all other variables in the problem, and \eqannref{b} follows by further requiring that $ \rv{W} $ is a Gaussian random variable and \autoref{lemma:thomas_lemma1}.
	In an identical manner, one can show that
	\begin{equation}
		I(\rv{X}_2;\rv{V}|\rv{U},\rv{X}_1) \leq  I(\rv{X}_2^{\mathcal{G}};\rv{V}^{\mathcal{G}}|\rv{U}^{\mathcal{G}},\rv{X}_1^{\mathcal{G}}).
	\end{equation}
	By shift-invariance of differential entropy there is no loss in optimality in restricting attention to zero-mean random variables. This completes the proof of the lemma.
\end{proof}

Equipped with \autoref{lemma:upper_bound_gaussian_function}, we proceed to evaluate \autoref{theorem:cran_2bs_upper_bound1} with the covariance matrix of the general form given by
\begin{equation}
	\Sigma \triangleq  
	\Exp{
		\begin{pmatrix}
			\rv{U} \\ \rv{X}_1\\ \rv{X}_2
		\end{pmatrix}	
		\begin{pmatrix}
			\rv{U} \\ \rv{X}_1\\ \rv{X}_2 
		\end{pmatrix}	^T
	}
	=
	\begin{pmatrix}
		P_u & \rho_1 \sqrt{P_u P_1} & \rho_2 \sqrt{P_u P_2}  \\
		\rho_1 \sqrt{P_u P_1} &P_1 &  \rho \sqrt{P_1 P_2} \\
		\rho_2 \sqrt{P_u P_2} & \rho \sqrt{P_1 P_2} & P_2 
	\end{pmatrix}.
\end{equation}
 We obtain the following upper bound on capacity.
\begin{proposition} \label{proposition:upper_bound1_gaussian}
	Rate $ R $ is achievable only if it satisfies the following constraints for some $ 0\leq \rho, \rho_1,\rho_2 \leq 1 $, such that $ 1 - \rho^2-\rho_1^2 - \rho_2^2+2\rho\rho_1\rho_2 \geq 0 $:
	\begin{equation} \label{eq:upper_bound_gaussian}
		R \leq \min_{N\geq 0} \min 
		\begin{cases}
			C_1+C_2 \\
			C_1+C_{12} +\frac{1}{2} \log \left(\frac{1- \rho_1^2 +  \left(1-\rho ^2-\rho_1^2-\rho_2^2+2 \rho  \rho_1 \rho_2 \right) P_2}{1-
				\rho_1^2}\right) \\
			C_2+C_{21} +\frac{1}{2} \log \left(\frac{1- \rho_2^2 +  \left(1-\rho ^2-\rho_1^2-\rho_2^2+2 \rho  \rho_1 \rho_2 \right) P_1}{1-
				\rho_2^2}\right) \\
			\frac{1}{2} \log \left(1 +P_1+P_2+  2 \rho  \sqrt{P_1 P_2}\right) \\
			C_{12} + C_{21}+\frac{1}{2} \log \left(  1+(1 - \rho_1^2) P_1+(1-\rho_2^2)P_2 + 2 (\rho - \rho_1 \rho_2)  \sqrt{P_1 P_2} \right),\\
			\frac{1}{2} \bigg[
			C_1+C_2+C_{12}+C_{21} 
			+ 
			\frac{1}{2} \log 
			\left(
			\frac{N
				\cdot 	\varphi(\rho,\rho_1,\rho_2,P_1,P_2)
				\phi_{1}(\rho,\rho_1,\rho_2,P_{2}) \cdot 
				\phi_{2}(\rho,\rho_1,\rho_2,P_{1})}{
				(1+N)^2(1-\rho_1^2)(1-\rho_2^2)	 
				\left[ N +  	\varphi(\rho,\rho_1,\rho_2,P_1,P_2)\right]
			}
			\right)
			\bigg]
			.
		\end{cases}
	\end{equation}
	where
	\begin{equation}
		\phi_{k}(\rho,\rho_1,\rho_2,P) \triangleq 
		(1-\rho_k)^2 (1+N) + (1-\rho^2-\rho_1^2-\rho_2^2 +2\rho \rho_1 \rho_2)P,
	\end{equation}
	and
	\begin{equation}
		\varphi(\rho,\rho_1,\rho_2,P_1,P_2) \triangleq 1+ (1-\rho_1^2)P_1+(1-\rho_2^2)P_2+2(\rho-\rho_1\rho_2)\sqrt{P_1 P_2}.
	\end{equation}
\end{proposition}

The above upper bound is hard to compute since it is a $ \min \max $ optimization problem. In a similar fashion to \cite{SaeediBidokhti2016}, we propose to choose a specific $ N $ that nullifies the term $ I(\rv{X}_1;\rv{X}_2|\rv{U},\rv{V}) $ in \eqref{eq:upper_bound_equivalent_term}. Note that for some values of $ (\rho,\rho_1,\rho_2,P_1,P_2) $ this $ N $ might be negative, in contradiction to $ \rv{W} $ being proper random variable, therefore, a more precise choice of $ N^* $ would be:
\begin{equation}
	N^* = \argmin_{N \geq 0} I(\rv{X}_1;\rv{X}_2|\rv{U},\rv{V}) .
\end{equation}
Thus, if $ N^* > 0 $, then an upper bound is given by
\begin{equation} \label{eq:upper_bound_gaussian2}
	R \leq \max_{(\rho,\rho_1,\rho_2)} \min 
	\begin{cases}
		C_1+C_2, \\
		C_1+C_{12} +\frac{1}{2} \log \left(\frac{1- \rho_1^2 +  \left(1-\rho ^2-\rho_1^2-\rho_2^2+2 \rho  \rho_1 \rho_2 \right) P_2}{1-
			\rho_1^2}\right), \\
		C_2+C_{21} +\frac{1}{2} \log \left(\frac{1- \rho_2^2 +  \left(1-\rho ^2-\rho_1^2-\rho_2^2+2 \rho  \rho_1 \rho_2 \right) P_1}{1-
			\rho_2^2}\right), \\
		\frac{1}{2} \log \left(1 +P_1+P_2+  2 \rho  \sqrt{P_1 P_2}\right), \\
		C_{12} + C_{21}+\frac{1}{2} \log \left(  1+(1 - \rho_1^2) P_1+(1-\rho_2^2)P_2 + 2 (\rho - \rho_1 \rho_2)  \sqrt{P_1 P_2} \right),\\
		\frac{1}{2} \left[ C_1 + C_2 + C_{12} + C_{21}+\frac{1}{2} \log \left(\frac{ \left( 1+(1 - \rho_1^2) P_1+(1-\rho_2^2)P_2 + 2 (\rho - \rho_1 \rho_2)  \sqrt{P_1 P_2} \right) \left(1-\rho ^2-\rho_1^2-\rho_2^2+2 \rho  \rho_1 \rho_2\right) }{\left(1-\rho_1^2\right) \left(1-\rho_2^2\right)} \right) \right]
		,
	\end{cases}
\end{equation}
for  $ 1 - \rho^2-\rho_1^2 - \rho_2^2+2\rho\rho_1\rho_2 \geq 0 $.
Otherwise, if $ N^* = 0 $, then $ \rv{V} = \rv{Y} $ and
\begin{equation} \label{eq:upper_bound_gaussian3}
	R \leq \max_{(\rho,\rho_1,\rho_2)} \min 
	\begin{cases}
		C_1+C_2, \\
		C_1+C_{12} +\frac{1}{2} \log \left(\frac{1- \rho_1^2 +  \left(1-\rho ^2-\rho_1^2-\rho_2^2+2 \rho  \rho_1 \rho_2 \right) P_2}{1-
			\rho_1^2}\right), \\
		C_2+C_{21} +\frac{1}{2} \log \left(\frac{1- \rho_2^2 +  \left(1-\rho ^2-\rho_1^2-\rho_2^2+2 \rho  \rho_1 \rho_2 \right) P_1}{1-
			\rho_2^2}\right), \\
		\frac{1}{2} \log \left(1 +P_1+P_2+  2 \rho  \sqrt{P_1 P_2}\right), \\
		C_{12} + C_{21}+\frac{1}{2} \log \left(  1+(1 - \rho_1^2) P_1+(1-\rho_2^2)P_2 + 2 (\rho - \rho_1 \rho_2)  \sqrt{P_1 P_2} \right)
		.
	\end{cases}
\end{equation}
\autoref{proposition:cran_2bs_gaussian_upper_bound1} then follows by taking $\rho_1 = \rho_2 = \rho_C$ and $P_1 = P_2 = P$ for the symmetric setting.

We proceed to evaluate our second upper bound from \autoref{theorem:cran_2bs_upper_bound2}.
\begin{proposition}\label{proposition:upper_bound2_gaussian}
	Rate $ R $ is achievable only if it satisfies the following constraints for some $ 0\leq \rho, \rho_1,\rho_2 \leq 1 $, such that $ 1 - \rho^2-\rho_1^2 - \rho_2^2+2\rho\rho_1\rho_2 \geq 0 $:
	\begin{equation} \label{eq:upper_bound_gaussian5}
		R \leq \min_{N\geq 0} \min 
		\begin{cases}
			C_1+C_2 \\
			C_1+C_{12} +\frac{1}{2} \log \left(\frac{1- \rho_1^2 +  \left(1-\rho ^2-\rho_1^2-\rho_2^2+2 \rho  \rho_1 \rho_2 \right) P_2}{1-
				\rho_1^2}\right) \\
			C_2+C_{21} +\frac{1}{2} \log \left(\frac{1- \rho_2^2 +  \left(1-\rho ^2-\rho_1^2-\rho_2^2+2 \rho  \rho_1 \rho_2 \right) P_1}{1-
				\rho_2^2}\right) \\
			\frac{1}{2} \log \left(1 +P_1+P_2+  2 \rho  \sqrt{P_1 P_2}\right) \\
			C_{12} + C_{21}+\frac{1}{2} \log \left(  1+(1 - \rho_1^2) P_1+(1-\rho_2^2)P_2 + 2 (\rho - \rho_1 \rho_2)  \sqrt{P_1 P_2} \right),\\
			C_1+C_2 - \frac{1}{2} \log \left(2^{2(R-C_{12}-C_{21})} + N \right) - \frac{1}{2} \log (1+N)
			+ 
			\frac{1}{2} \log 
			\left(
			\frac{
				\phi_{1}(\rho,\rho_1,\rho_2,P_{2}) \cdot 
				\phi_{2}(\rho,\rho_1,\rho_2,P_{1})}{
				(1-\rho_1^2)(1-\rho_2^2)	
			}
			\right)
			.
		\end{cases}
	\end{equation}
	where
	\begin{equation}
		\phi_{k}(\rho,\rho_1,\rho_2,P) \triangleq 
		(1-\rho_k^2) (1+N) + (1-\rho^2-\rho_1^2-\rho_2^2 +2\rho \rho_1 \rho_2)P.
	\end{equation}
\end{proposition}
\begin{proof}
	The first 5 terms in \eqref{eq:upper_bound2x} may be loosened by dropping the random variable $ \rv{T} $, and therefore are identical to the ones in \autoref{proposition:upper_bound1_gaussian}. We would like to evaluate and bound the last constraint in \autoref{theorem:cran_2bs_upper_bound2}.
	\begin{align}
		R &\leq C_1+C_2 -I(\rv{X}_1,\rv{X}_2;\rv{V}|\rv{U},\rv{T}) + I(\rv{X}_1;\rv{V}|\rv{X}_2,\rv{U},\rv{T}) \\
		&+ I(\rv{X}_2;\rv{V}|\rv{X}_1,\rv{U},\rv{T}) \\
		&= C_1+C_2 -h(\rv{V}|\rv{U},\rv{T}) - h(\rv{V}|\rv{X}_1,\rv{X}_2) \\
		&+ h(\rv{V}|\rv{U},\rv{X}_1,\rv{T}) + h(\rv{V}|\rv{U},\rv{X}_2,\rv{T}) \\
		&\leq C_1+C_2 -h(\rv{V}|\rv{U},\rv{T}) - h(\rv{V}|\rv{X}_1,\rv{X}_2) \\
		& + h(\rv{V}|\rv{U},\rv{X}_1) + h(\rv{V}|\rv{U},\rv{X}_2).
	\end{align}
	Consider the term $ h(\rv{V}|\rv{U},\rv{T}) $. Since $ \rv{V} = \rv{Y} + \rv{W} $, with $ \rv{W} \sim \mathcal{N}(0,N) $, using the conditional EPI \cite[p. 22]{Gamal2011} we have the following lower bound on this term,
	\begin{equation}
		h(\rv{V}|\rv{U},\rv{T}) 
		\geq \frac{1}{2} \log \left(2^{2h(\rv{Y}|\rv{U},\rv{T})} +2\pi e N \right).
	\end{equation}
	Furthermore, rearranging the fifth constraint from \eqref{eq:upper_bound2x}, we obtain the following inequality
	\begin{align}
		h(\rv{Y}|\rv{U},\rv{T})  
		&= I(\rv{X}_1,\rv{X}_2;\rv{Y}|\rv{U},\rv{T}) + \frac{1}{2} \log (2\pi e) \\
		&\geq R -C_{12}- C_{21} + \frac{1}{2} \log (2\pi e).
	\end{align}
	Thus,
	\begin{equation}
		h(\rv{V}|\rv{U},\rv{T}) 
		\geq \frac{1}{2} \log 2\pi e \left(2^{2(R-C_{12}-C_{21})} + N \right).
	\end{equation}
	We conclude that
	\begin{multline}
		R \leq C_1+C_2 - \frac{1}{2} \log \left(2^{2(R-C_{12}-C_{21})} + N \right) - \frac{1}{2} \log (1+N)
		\\
		+ \frac{1}{2} \log \frac{(1-\rho_2^2)(1+N) +(1-\rho^2 - \rho_1^2 - \rho_2^2 +2\rho \rho_1 \rho_2)P_1}{1-\rho_2^2} \\
		+ \frac{1}{2} \log \frac{(1-\rho_1^2)(1+N) +(1-\rho^2 - \rho_1^2 - \rho_2^2 +2\rho \rho_1 \rho_2)P_2}{1-\rho_1^2}.
	\end{multline}
\end{proof}

\begin{remark}
	The last constraint in \eqref{eq:upper_bound_gaussian5} can be reformulated as:
		\begin{equation} \label{eq:cran_2bs_gaussian_upper_bound1}
		R \leq \frac{1}{2} \log \left(
		\sqrt{ 2^{2(C_1+C_2+C_{12}+C_{21})}  \frac{
				\phi_{1}(\rho,\rho_1,\rho_2,P_{2}) \cdot 
				\phi_{2}(\rho,\rho_1,\rho_2,P_{1})}{
				(1-\rho_1^2)(1-\rho_2^2)(1+N)
			} + \frac{2^{4(C_{12}+C_{21})} N^2}{4} } - \frac{2^{2(C_{12}+C_{21})} N }{2}
		\right)
	\end{equation}
	
\end{remark}
\begin{proof}
	Taking both sided of \eqref{eq:upper_bound_gaussian5} to the the power $ 2^{2 \cdot} $, results in the following inequality
	\begin{multline}
		2^{2(R-C_1-C_2)} \cdot \left(2^{2(R-C_{12}-C_{21})} + N \right) \\
		\leq 
		\frac{
			\phi_{1}(\rho,\rho_1,\rho_2,P_{2}) \cdot 
			\phi_{2}(\rho,\rho_1,\rho_2,P_{1})}{(1+N)
			(1-\rho_1^2)(1-\rho_2^2)	
		}
	\end{multline}
	Rearranging terms and denoting $ \xi = 2^{2R} $, we obtain the following convex set
	\begin{multline}
		\xi^2 + 2^{2(C_{12}+C_{21})} N \xi \\
		 \leq  2^{2(C_1+C_2+C_{12}+C_{21})} \frac{
			\phi_{1}(\rho,\rho_1,\rho_2,P_{2}) \cdot 
			\phi_{2}(\rho,\rho_1,\rho_2,P_{1})}{
			(1+N)(1-\rho_1^2)(1-\rho_2^2)	
		},
	\end{multline}
	which implies
	\begin{equation}
		\xi \leq \frac{\sqrt{2^{4(C_{12}+C_{21})} N^2+4\cdot 2^{2(C_1+C_2+C_{12}+C_{21})}  \frac{
					\phi_{1}(\rho,\rho_1,\rho_2,P_{2}) \cdot 
					\phi_{2}(\rho,\rho_1,\rho_2,P_{1})}{
					(1+N)(1-\rho_1^2)(1-\rho_2^2)	
			} } - 2^{2(C_{12}+C_{21})} N }{2}.
	\end{equation}
	Taking $ R = \frac{1}{2} \log \xi $ completes the proof.
\end{proof}

Finally, \autoref{proposition:cran_2bs_gaussian_upper_bound2} follows by setting $\rho_1 = \rho_2 = \rho_C$ and $P_1 = P_2 = P$.


%
	
	\subsection{Coding Scheme} \label{appendix:cran_3bs_coding_scheme}
	\paragraph*{Codebook Generation}

Let $ \Omega = \{0,1,2,3\} $. Fix a joint \acrshort{pmf} $ \mathsf{P}_{\rv{U}_{\Omega} }  $, a set of rates $ \{ R_{\omega} \} $, $ \omega \in \Omega $, additional set of rates $ \{ R_{\omega \omega'} \} $, $ (\omega, \omega') \in \Omega \times \Omega-\{(1,0),(2,0),(3,0) \} $ and functions $ x_1(u_0,u_1) $, $ x_2(u_0, u_{2}) $, and $ x_3(u_0,u_3) $. Randomly and independently generate sequences $ u_\omega^n(k_\omega) $,  each according to $ \prod_{i=1}^n \mathsf{P}_{\rv{U}_\omega} (u_{\omega, i}) $, for $ k_\omega \in \setrn{R_{\omega}}{n}  $; for every $ \omega \in \Omega $.

Next, for every $ k_0 \in \setrn{R_{0}}{n} $, we generate the following three dictionaries, consisting of sequences that are conditionally typical given $ u_0^{n}(k_0) $,
\begin{equation}
	\mathcal{D}_1(k_0) \triangleq \{ k_1  \in \setrn{R_{1}}{n} : u_1^n(k_1)  \in \typsetn{\epsilon'}{\rv{U}_1|u_0^n(k_0) }  \},
\end{equation}
\begin{equation}
	\mathcal{D}_2(k_0) \triangleq \{ k_2  \in \setrn{R_{2}}{n} : u_2^n(k_2)  \in \typsetn{\epsilon'}{\rv{U}_2|u_0^n(k_0) }  \},
\end{equation}
\begin{equation}
	\mathcal{D}_3(k_0) \triangleq \{ k_3  \in \setrn{R_{3}}{n} : u_3^n(k_3)  \in \typsetn{\epsilon'}{\rv{U}_3|u_0^n(k_0) }  \}.
\end{equation}

Every index tuple in the dictionaries is assigned a unique reference label by means of the functions
\begin{equation}
	\delta_1(\cdot|k_0): \mathcal{D}_1(k_0) \rightarrow \{1,\dots, |\set{D}_1(k_0) |\},
\end{equation}
\begin{equation}
	\delta_2(\cdot|k_0): \mathcal{D}_2(k_0) \rightarrow \{1,\dots, |\set{D}_2(k_0) |\},
\end{equation}
\begin{equation}
	\delta_3(\cdot|k_0): \mathcal{D}_3(k_0) \rightarrow \{1,\dots, |\set{D}_3(k_0) |\}.
\end{equation}
Let $ \delta_1^{-1}(\cdot|k_0) $, $ \delta_2^{-1}(\cdot|k_0) $, and $ \delta_3^{-1}(\cdot|k_0) $ denote the corresponding inverse maps.

Finally, we randomly and independently assign an index $ m(k_{\Omega}) $ to each index tuple $ k_{\Omega} \in \prod_{\omega \in\Omega} \setrn{R_\omega}{n}  $ according to a uniform \acrshort{pmf} over $ \setrn{R}{n} $. We refer to each subset of index tuples with the same index $ m $ as a bin $ \mathcal{B}(m) $.

\paragraph*{Central Processor} Fix $ \epsilon' > 0 $. Upon observing $ m  $, the \acrshort{cp} finds $ k_{\Omega} \in \set{B}(m)  $ such that
\begin{equation}
	u_{\Omega}^n(k_{\Omega}) \in \mathcal{T}_{\epsilon'}^{(n)} (\rv{U}_{\Omega}).
\end{equation}
If there is more than one such tuple, choose an arbitrary one among them. If no such tuple exists, choose $ k_{\Omega} = 1_{\Omega} $. 
Then, the \acrshort{cp} splits $ k_0 $ into four subindices $ m_{00} $, $ m_{01} $, $ m_{02} $ and $ m_{03} $ of rates $ R_{00} $, $ R_{01} $, $ R_{02} $, and $ R_{03} $ respectively. 

In addition, the \acrshort{cp} splits $ \delta_1(k_1|k_0) $ into three subindices $ m_{11} $, $ m_{12} $ and $ m_{13} $ of rates $ R_{11} $, $ R_{12} $ and $ R_{13} $, respectively. Similarly, it  splits $ \delta_2(k_2|k_0) $ into three subindices $ m_{21} $, $ m_{22} $ and $ m_{23} $ of rates $ R_{21} $, $ R_{22} $ and $ R_{23} $, respectively. It also  splits $ \delta_3(k_3|k_0) $ into three subindices $ m_{31} $, $ m_{32} $ and $ m_{33} $ of rates $ R_{31} $, $ R_{32} $ and $ R_{33} $, respectively. 

Finally, the \acrshort{cp} sends the index tuple $ (m_{00}, m_{01}, m_{11},m_{21},m_{31}) $ to BS1, $ (m_{00}, m_{02}, m_{12},m_{22}, m_{32}) $ to BS 2, and $ (m_{00}, m_{03},\allowbreak m_{13},m_{23},  m_{33}) $ to BS 3.
The encoding operation at the \acrshort{cp} is illustrated in \autoref{figure:3bs_1ms_scc_encoding}.

\paragraph*{Base Stations}
BS 1 forwards $ (m_{01}, m_{21}) $ to BS 2 over the cooperation link. BS 1 forwards $ (m_{01},m_{31}) $ to BS 3 over the cooperation link. BS 2 forwards $ (m_{02},m_{12}) $ to BS 1 over the cooperation link. BS 2 forwards $ (m_{02},m_{32}) $ to BS 3 over the cooperation link. BS 3 forwards $ (m_{03},m_{23}) $ to BS 2 over the cooperation link. BS 3 forwards $ (m_{03},m_{13}) $ to BS 1 over the cooperation link.

All BSs recover the index $ k_0 $ from the tuple $ (m_{00},m_{01}, m_{02}, m_{03}) $. 

Then BS 1 applies the reverse mapping $ \delta_1^{-1}(\cdot|k_0) $ to the obtained $ (m_{11},m_{12},m_{13}) $ to recover $ k_1 $. Similarly, BS 2 applies the reverse mapping $ \delta_2^{-1}(\cdot|k_0) $ to the obtained $ (m_{21},m_{22},m_{23}) $ to recover $ k_2 $. In addition, BS 3 applies the reverse mapping $ \delta_3^{-1}(\cdot|k_0) $ to the obtained $ (m_{31},m_{32},m_{33}) $ to recover $ k_3 $.

Finally, BS 1 transmits the symbol $ x_{1i} (u_{0,i}(k_0),u_{1,i}(k_1)) $, BS 2 transmits the symbol $ x_{2i} (u_{0,i}(k_0), u_{2,i}(k_2)) $, and BS 3 transmits the symbol $ x_{3i} (u_{0,i}(k_0), u_{3,i}(k_3)) $, at each time $ i \in [1:n] $.

\paragraph*{Mobile User}
Let $ \epsilon > \epsilon' $. The User declares that $ \hat{m} $ is sent if it is the unique message such that for some $ k_{\Omega}  \in \mathcal{B}(\hat{m}) $ it holds that
\begin{equation}
	(u^n_{\Omega}(k_{\Omega}),y^n) \in \mathcal{T}_{\epsilon}^{(n)} (\rv{U}_{\Omega}, \rv{Y});
\end{equation}
otherwise it declares an error.

\paragraph*{Analysis of error probability:}
Let $ M $ be the message and let $ K_{\Omega} $ be the indices chosen at the encoder. In order to have a lossless transmission over the digital links, the following system of inequalities must hold:
\begin{align}
	R_{00} + R_{01} + R_{11} + R_{21} +  R_{31}  &\leq C_{11},\\
	R_{00} + R_{02} +  R_{12} + R_{22} + R_{32}   &\leq C_{22},\\
	R_{00} + R_{03} + R_{13} + R_{23} +  R_{33}  &\leq C_{33},\\
	R_{01} + R_{21} &\leq C_{21}, \\
	R_{02} + R_{12} &\leq C_{12}, \\
	R_{02} + R_{32} &\leq C_{32}, \\
	R_{03} + R_{23} &\leq C_{23}, \\
	R_{03} + R_{13} &\leq C_{13}, \\
	R_{01} + R_{31} &\leq C_{31}.
\end{align}
Also, we note that
\begin{align}
	R_{00} + R_{01} + R_{02} + R_{03} &= R_{0}, \\
	R_{11}+R_{12}+R_{13} &= \log |\mathcal{D}_1(K_0)|, \\
	R_{21}+R_{22}+R_{23} &= \log |\mathcal{D}_2(K_0)|, \\
	R_{31}+R_{32}+R_{33} &= \log |\mathcal{D}_3(K_0)|.
\end{align}
Thus, after applying Fourier-Motzkin elimination we obtain:
\begin{align}
	R_0 + \log |\mathcal{D}_1(K_0)|  &\leq C_{11} + C_{12} + C_{13}, \label{eq:scc_codebook_size1} \\
	R_0 + \log |\mathcal{D}_2(K_0)|  &\leq C_{21}+ C_{22} + C_{23}, \label{eq:scc_codebook_size2}\\
	R_0 + \log |\mathcal{D}_3(K_0)|  &\leq C_{31} + C_{32} + C_{33}, \label{eq:scc_codebook_size3}\\
	R_0 + \log |\mathcal{D}_1(K_0)|+ \log |\mathcal{D}_2(K_0)|+ \log |\mathcal{D}_3(K_0)|  & \leq C_{11}+ C_{22} + C_{33} . \label{eq:scc_codebook_size4}
\end{align}
We denote by $ \mathcal{A} $ the intersection of the random events \eqref{eq:scc_codebook_size1}--\eqref{eq:scc_codebook_size4}. Applying \autoref{lemma:expected_size_of_codebooks} proved in Appendix \ref{section:expected_size_of_codebooks}, the random event $ \mathcal{A} $ happens with high probability as $ n\rightarrow \infty $ if
\begin{align}
	C_{11} + C_{12} + C_{13} &\geq R_0 + R_1  -I(\rv{U}_0;\rv{U}_1) , \\
	C_{21} + C_{22} + C_{23} &\geq R_0 + R_2  -I(\rv{U}_0;\rv{U}_2) , \\
	C_{31} + C_{32} +C_{33} &\geq R_0 + R_3  -I(\rv{U}_0;\rv{U}_3) , \\
	C_{11}+C_{22} + C_{33} & \geq R_0 + R_1 + R_2 + R_3 -I(\rv{U}_0;\rv{U}_1)  -I(\rv{U}_0;\rv{U}_2)  -I(\rv{U}_0;\rv{U}_3).
\end{align}
In addition to the error event $ \mathcal{A}^c $, the decoding fails if one or more of the following events occur:
\begin{align}
	\mathcal{E}_{CP} &= \{ \rv{U}_{\Omega} \notin \typsetn{\epsilon'}{\rv{U}_{\Omega}}  \text{ for all } k_{\Omega} \in \mathcal{B}(M) \}, \\
	\mathcal{E}_{d_0} &= \{ (\rv{U}_{\Omega}^n(K_{\Omega}), \rv{Y}^n) \notin \typsetn{\epsilon}{\rv{U}_{\Omega} , \rv{Y}}  \} ,
\end{align}
and for every nonempty $ \set{S} \subseteq \Omega $,
\begin{equation}
	\mathcal{E}_{d_\set{S}} = \{(\rv{U}_{\set{S}}^n(k_{\set{S}}), \rv{U}_{\set{S}^c}^n(K_{\set{S}^c}), \rv{Y}^n) \in \mathcal{T}_{\epsilon}^{(n)} (\mathsf{P}_{\rv{U}_{\Omega} \rv{Y}}) \text{ for some } k_{\set{S}}\neq K_{\set{S}}  \} .
\end{equation}

Thus, the average probability of error is upper bounded as
\begin{align}
	P_e^{(n)}  &= \Prob{\hat{\rv{M}} \neq \rv{M}} \\
	&\leq \Prob{\mathcal{E}_{CP}} + \Prob{\mathcal{A}^c} + \Prob{\mathcal{E}_{d_0} \cap \mathcal{E}_{CP}^c \cap \mathcal{A}} + \sum_{\set{S}\subseteq \Omega} \Prob{\mathcal{E}_{d_\set{S}}}.
\end{align}
Applying \autoref{lemma:covering_lemma} proved in Appendix \ref{section:covering_lemma}, the term $ \Prob{\mathcal{E}_{CP}}  $ tends to zero as $ n \rightarrow \infty $ if
\begin{equation}
	\sum_{\omega \in \set{S}} R_{\omega} > \indicator \{\set{S} = \Omega\} \cdot  R + \Gamma(\rv{U}_{\set{S}}) ,
\end{equation}
for all $ \set{S} \subseteq \Omega  $, such that $ |\set{S}| \geq 2 $.

Next, due to codebook construction and the conditional typicality lemma, $  \Prob{\mathcal{E}_1 \cap \mathcal{E}_{CP}^c \cap \mathcal{A}} $ tends to zero as $ n \rightarrow \infty $. From \autoref{lemma:packing_lemma} proved in Appendix \ref{section:packing_lemma}, $ \Prob{\mathcal{E}_{d_{\set{S}}}} $ tends to zero as $ n \rightarrow \infty $ if
\begin{equation}
	\sum_{\omega \in \set{S}} R_{\omega} < I(\rv{U}_{\set{S}}; \rv{U}_{\set{S}^c},\rv{Y}) + \Gamma (\rv{U}_{\set{S}}).
\end{equation}
Finally, the theorem is established by letting $ \epsilon $ tend to zero.

	\subsection{Proof of \autoref{theorem:cran_3bs_upper_bound2}} \label{appendix:upper_bound2_proof}
	From \eqref{eq:multiletter_upper_bound1} we have
\begin{multline}
	nR \leq n\sum_{i=1}^{3} C_{ii} 
	- 2 I(\rv{X}_1^n,\rv{X}_2^n,\rv{X}_3^n;\rv{V}_n| \rv{W}_c) \\
	+ I(\rv{V}^n;\rv{X}_1^n,\rv{X}_2^n|\rv{X}_3^n| \rv{W}_c)
	+I(\rv{V}^n;\rv{X}_1^n,\rv{X}_3^n|\rv{X}_2^n| \rv{W}_c) \\
	+I(\rv{V}^n;\rv{X}_2^n,\rv{X}_3^n|\rv{X}_1^n| \rv{W}_c).
\end{multline}

As was suggested in \cite[Proof of Theorem 3]{SaeediBidokhti2016}, since $ \rv{V}^n $ is arbitrary, we choose each symbol $ 
\rv{V}_i $ as the output of the channel  $ \mathsf{P}_{\rv{V}|\rv{Y}}(v_i|y_i) $, with input $ y_i $, $ i=1,2,\dots,n $. 
Furthermore, we denote $ \tilde{\rv{U}}_i = \rv{W}_{c} $. With this choice of $ \rv{V}_i $ and $ \tilde{\rv{U}}_i $, we can expand $ I(\rv{X}_1^n,\rv{X}_2^n,\rv{X}_3^n;\rv{V}^n| \rv{W}_c)  $ as follows:
\begin{align}
	I(\rv{X}_{[3]}^n;\rv{V}^n| \rv{W}_c) 
	&= \sum_{i=1}^n I(\rv{X}_{[3]}^n;\rv{V}_i|\rv{W}_{c},\rv{V}^{i-1}) \\
	&= \sum_{i=1}^n I(\rv{X}_{{[3]},i};\rv{V}_i|\tilde{\rv{U}}_{i},\rv{V}^{i-1}) \\
	&= n I(\rv{X}_{{[3]},\rv{Q}};\rv{V}_{\rv{Q}}|\tilde{\rv{U}}_{\rv{Q}},\rv{V}^{\rv{Q}-1},\rv{Q}).
\end{align}
Similarly,
\begin{align}
	I(\rv{X}_{2}^n,\rv{X}_3^n;\rv{V}^n|\rv{W}_{12},\rv{W}_{21},\rv{X_1^n})  
	&= \sum_{i=1}^n I(\rv{X}_{2}^n,\rv{X}_3^n;\rv{V}_i|\rv{W}_{12},\rv{W}_{21},\rv{X_1^n},\rv{V}^{i-1})  \\
	&\leq \sum_{i=1}^n I(\rv{X}_{2i},\rv{X}_{3i};\rv{V}_i|\tilde{\rv{U}}_i,\rv{X_{1i}},\rv{V}^{i-1})  \\
	&\leq n I(\rv{X}_{2\rv{Q}}, \rv{X}_{3\rv{Q}};\rv{V}_{\rv{Q}}|\tilde{\rv{U}}_{\rv{Q}},\rv{X_{1\rv{Q}}},\rv{V}^{\rv{Q}-1},\rv{Q}).
\end{align}
Also,
\begin{equation}
	I(\rv{X}_1^n, \rv{X}_3^n;\rv{V}^n|\rv{W}_{c},\rv{X}_2^n)
	\leq
	n I(\rv{X}_{1\rv{Q}},\rv{X}_{3\rv{Q}};\rv{V}_{\rv{Q}}|\tilde{\rv{U}}_{\rv{Q}},\rv{X_{2\rv{Q}}},\rv{V}^{\rv{Q}-1},\rv{Q}).
\end{equation}
and
\begin{equation}
	I(\rv{X}_1^n, \rv{X}_2^n;\rv{V}^n|\rv{W}_{c},\rv{X}_2^n)
	\leq
	n I(\rv{X}_{1\rv{Q}},\rv{X}_{2\rv{Q}};\rv{V}_{\rv{Q}}|\tilde{\rv{U}}_{\rv{Q}},\rv{X_{1\rv{Q}}},\rv{V}^{\rv{Q}-1},\rv{Q}).
\end{equation}
Combining the above inequalities we obtain:
\begin{multline}
	R \leq \sum_{i=1}^{3} C_{ii} 
	- 2I(\rv{X}_{\Omega, \rv{Q}};\rv{V}_{\rv{Q}}|\tilde{\rv{U}}_{\rv{Q}},\rv{V}^{\rv{Q}-1},\rv{Q}) \\
	+ I(\rv{X}_{1\rv{Q}}, \rv{X}_{3\rv{Q}};\rv{V}_{\rv{Q}}|\tilde{\rv{U}}_{\rv{Q}},\rv{V}^{\rv{Q}-1},\rv{X}_{2\rv{Q}},\rv{Q}) \\
	+ I(\rv{X}_{2\rv{Q}},\rv{X}_{3\rv{Q}};\rv{V}_{\rv{Q}}|\tilde{\rv{U}}_{\rv{Q}},\rv{V}^{\rv{Q}-1},\rv{X}_{1\rv{Q}},\rv{Q})  \\
	+ I(\rv{X}_{1\rv{Q}},\rv{X}_{2\rv{Q}};\rv{V}_{\rv{Q}}|\tilde{\rv{U}}_{\rv{Q}},\rv{V}^{\rv{Q}-1},\rv{X}_{3\rv{Q}},\rv{Q}) .
\end{multline}

Next, consider the multi-letter mutual information term in the \acrshort{rhs} of \eqref{eq:multiletter_upper_bound2},
\begin{align}
	 I(\rv{X}_{\set{S}^c}^n;\rv{Y}^n|\rv{W}_{c},\rv{X}_{\set{S}}^n)
	 &= \sum_{i=1}^n  I(\rv{X}_{\set{S}^c}^n;\rv{Y}_i|\rv{W}_{c},\rv{X}_{\set{S}}^n,\rv{Y}^{i-1}) \\
	 &\leq \sum_{i=1}^n  I(\rv{X}_{\set{S}^c,i};\rv{Y}_i|\tilde{\rv{U}}_{i},\rv{X}_{\set{S}.i},\rv{Y}^{i-1}) \\
	 &\eqann{=}{a} \sum_{i=1}^n  I(\rv{X}_{\set{S}^c,i};\rv{Y}_i|\tilde{\rv{U}}_{i},\rv{X}_{\set{S},i},\rv{Y}^{i-1},\rv{V}^{i-1}) \\
	 & \leq \sum_{i=1}^n  I(\rv{X}_{\set{S}^c,i};\rv{Y}_i|\tilde{\rv{U}}_{i},\rv{X}_{\set{S},i},\rv{V}^{i-1}) \\
	 & = n  I(\rv{X}_{\set{S}^c, \rv{Q}};\rv{Y}_{\rv{Q}}|\tilde{\rv{U}}_{\rv{Q}},\rv{X}_{\set{S},\rv{Q}},\rv{V}^{\rv{Q}-1},\rv{Q}) ,
\end{align}
where \eqannref{a} follows since $ \rv{V}^n $ is the output of $ \prod_{i=1}^n \mathsf{P}_{\rv{V}|\rv{Y}}(v_i|y_i)  $.


To this end we obtained the following list of inequalities:
\begin{align}
	R & \leq \sum_{i=1}^{3} C_{ii} 
	- 2I(\rv{X}_{\Omega, \rv{Q}};\rv{V}_{\rv{Q}}|\tilde{\rv{U}}_{\rv{Q}},\rv{V}^{\rv{Q}-1},\rv{Q}) \\ 
	&+ I(\rv{X}_{1\rv{Q}}, \rv{X}_{3\rv{Q}};\rv{V}_{\rv{Q}}|\tilde{\rv{U}}_{\rv{Q}},\rv{V}^{\rv{Q}-1},\rv{X}_{2\rv{Q}},\rv{Q}) \\
	&+ I(\rv{X}_{2\rv{Q}},\rv{X}_{3\rv{Q}};\rv{V}_{\rv{Q}}|\tilde{\rv{U}}_{\rv{Q}},\rv{V}^{\rv{Q}-1},\rv{X}_{1\rv{Q}},\rv{Q})  \\
	& + I(\rv{X}_{1\rv{Q}},\rv{X}_{2\rv{Q}};\rv{V}_{\rv{Q}}|\tilde{\rv{U}}_{\rv{Q}},\rv{V}^{\rv{Q}-1},\rv{X}_{3\rv{Q}},\rv{Q}) \\
	R & \leq \sum_{ \omega \in \set{S}} C_{\omega} + \sum_{\omega \in [3]}  \sum_{\omega' \neq \omega \in \set{S}^c} C_{\omega,\omega'} \\ 
	&+   I(\rv{X}_{\set{S}^c, \rv{Q}};\rv{Y}_{\rv{Q}}|\tilde{\rv{U}}_{\rv{Q}},\rv{X}_{\set{S},\rv{Q}},\rv{V}^{\rv{Q}-1},\rv{Q})\\
	R & \leq I(\rv{X}_{[3], \rv{Q}};\rv{Y}_{\rv{Q}}|\rv{Q}) 
\end{align}
Denoting $ \rv{X}_1 = \rv{X}_{1\rv{Q}} $, $ \rv{X}_2 = \rv{X}_{2\rv{Q}} $, $ \rv{Y} = \rv{Y}_{\rv{Q}} $, $ \rv{V} = \rv{V}_{\rv{Q}} $,   $ \tilde{\rv{U}} = \tilde{\rv{U}}_{\rv{Q}} $ and $ \rv{T} = \rv{V}^{\rv{Q}-1} $.
Furthermore, since $ I(\rv{X}_{[3]};\rv{Y}|\rv{Q}) \leq I(\rv{Q},\rv{X}_{[3]};\rv{Y})  = I(\rv{X}_{[3]};\rv{Y})   $, identifying $ \rv{U} = (\tilde{\rv{U}},\rv{Q}) $  completes the proof of the theorem.

%

	
	\subsection{Lower Bound Proof for the 3 relay Gaussian channel}
    \label{appendix:cran_3bs_gaussian_lower_bound_proof}
	Now let us specify our (sub-optimal) choice of auxiliary random variables for the various schemes. Let $ \Sigma_{\rv{U} \rv{X}_1   \rv{X}_2 \rv{X}_3} $ a general covariance matrix for the triple $ (\rv{U},\rv{X}_1,\rv{X}_2,\rv{X}_3) $ of the form
\begin{equation} \label{eq:cran_3bs1mu_general_covariance_matrix}
	\Sigma_{\rv{U} \rv{X}_1   \rv{X}_2 \rv{X}_3} =
	\begin{pmatrix}
		P_u & \rho_1 \sqrt{P_u P_1} & \rho_2 \sqrt{P_u P_2} & \rho_3 \sqrt{P_u P_3} \\
		\rho_1 \sqrt{P_u P_1} & P_1 & \rho_{12} \sqrt{P_1 P_2} & \rho_{13} \sqrt{P_1 P_3} \\
		\rho_2 \sqrt{P_u P_2} & \rho_{12} \sqrt{P_1 P_2} & P_2  & \rho_{23} \sqrt{P_2 P_3} \\
		\rho_3 \sqrt{P_u P_3} & \rho_{13} \sqrt{P_1 P_3}  & \rho_{23} \sqrt{P_2 P_3} & P_3
	\end{pmatrix}.
\end{equation}
For the symmetric case we may restrict the covariance matrix over which optimization takes place, i.e, \eqref{eq:cran_3bs1mu_general_covariance_matrix}, to $ \rho_1 = \rho_2 = \rho_3 = \rho_c $ and $\rho = \rho_{12} = \rho_{13} = \rho_{23}$. 
With this choice, we evaluate \autoref{corollary:computable_lower_bound} and obtain the desired bound in \autoref{proposition:cran_3bs_gaussian_lower_bound}.

	\subsection{Upper Bound Proof for the 3 relay Gaussian channel}
	\label{appendix:cran_3bs_gaussian_upper_bound2_proof}
	In this section we will specialize \autoref{theorem:cran_3bs_upper_bound2} for the symmetric Gaussian setting.
The second term in \eqref{eq:upper_bound2_3relays_noQ} may be loosened by dropping the random variable $ \rv{T} $, and therefore is maximized by symmetric jointly Gaussian $(\rv{U},\rv{X}_{[3]},\rv{Y},\rv{T})$, in a similar manner as we have for \autoref{proposition:cran_3bs_gaussian_lower_bound}. We would like to evaluate and bound the last constraint in \autoref{theorem:cran_3bs_upper_bound2}.
\begin{align*}
	R &\leq \sum_{i=1}^{3} C_{ii} \minus  2I(\rv{X}_{[3]};\rv{V}|\rv{U},\rv{T}) + I(\rv{X}_{1}, \rv{X}_{2};\rv{V}|\rv{U},\rv{T},\rv{X}_{3}) \\
	&+ I(\rv{X}_{1}, \rv{X}_{3};\rv{V}|\rv{U},\rv{T},\rv{X}_{2})+ I(\rv{X}_{2}, \rv{X}_{3};\rv{V}|\rv{U},\rv{T},\rv{X}_{1}) \\
	&= \sum_{i=1}^{3} C_{ii} \minus  2h(\rv{V}|\rv{U},\rv{T}) + h(\rv{V}|\rv{U},\rv{T},\rv{X}_{3}) \\
	&+ h(\rv{V}|\rv{U},\rv{T},\rv{X}_{2})+ h(\rv{V}|\rv{U},\rv{T},\rv{X}_{1}) + h(\rv{V}|\rv{X}_{[3]}) .
\end{align*}
Consider the term $ h(\rv{V}|\rv{U},\rv{T}) $. Since $ \rv{V} = \rv{Y} + \rv{W} $, with $ \rv{W} \sim \mathcal{N}(0,N) $, using the conditional EPI \cite[p. 22]{Gamal2011} we have the following lower bound on this term,
\begin{equation}
	h(\rv{V}|\rv{U},\rv{T}) 
	\geq \frac{1}{2} \log \left(2^{2h(\rv{Y}|\rv{U},\rv{T})} +2\pi e N \right).
\end{equation}
Furthermore,
\begin{align}
	h(\rv{Y}|\rv{U},\rv{T})  
	&= I(\rv{X}_{[3]};\rv{Y}|\rv{U},\rv{T}) + \frac{1}{2} \log (2\pi e) \\
	& \geq R - \sum_{
		\substack{\omega \in [3] \\ \omega' \neq \omega \in \set{S}^c}
	} C_{\omega \omega'} 
+ \frac{1}{2} \log (2\pi e).
\end{align}
Thus,
\begin{equation}
	h(\rv{V}|\rv{U},\rv{T}) 
	\geq \frac{1}{2} \log 2\pi e \left(2^{2(R- 6 C_0 )} + N \right).
\end{equation}
Furthermore,
\begin{align}
	h(\rv{V}|\rv{U},\rv{T},\rv{X}_{1})
	&\leq h(\rv{V}|\rv{U},\rv{X}_{1}) \\
	&\leq \frac{1}{2} \log \left(2\pi e\right).
\end{align}
Due symmetry of the problem, we claim that the optimal covariance matrix of $ (\rv{U},\rv{X}_1,\rv{X}_2,\rv{X}_2) $ has a super-symmetric form, i.e.,
\begin{equation}
	\Sigma \triangleq  
	\Exp{
		\begin{pmatrix}
			\rv{U} \\ \rv{X}_1\\ \rv{X}_2 \\ \rv{X}_3\\
		\end{pmatrix}	
		\begin{pmatrix}
			\rv{U} \\ \rv{X}_1\\ \rv{X}_2 \\ \rv{X}_3\\
		\end{pmatrix}	^T
	}
	=
	\begin{pmatrix}
		P_u & \rho_c \sqrt{P_u P} & \rho_c \sqrt{P_u P} & \rho_c \sqrt{P_u P} \\
		\rho_c \sqrt{P_u P} &P & P \rho &  P \rho\\
		\rho_c \sqrt{P_u P} &P \rho & P & P \rho  \\
		\rho_c \sqrt{P_u P} &P \rho & P \rho & P
	\end{pmatrix}.
\end{equation}
We conclude that
\begin{multline}
	R \leq 3C - \frac{1}{2} \log \left(2^{2(R-6C_{0})} + N \right) - \frac{1}{2} \log (1+N)
	\\
	+ \frac{3}{2} \log \frac{(1-\rho_c^2)(1+N) +2 (1 + \rho -2\rho^2 - 3\rho_c^2  +3\rho \rho_c^2)P}{(1-\rho_c^2)} .
\end{multline}

    \subsection{Proof of \autoref{lemma:dpi_cran_conf}} \label{appendix:dpi_cran_conf_proof}
	Denote $ \rv{W}_c \triangleq (\rv{W}_{12},\rv{W}_{13},\rv{W}_{21},\rv{W}_{23},\rv{W}_{31},\rv{W}_{32}) $. First, we use the chain rule to expand $ \Gamma(\rv{X}_1^n; \rv{X}_2^n; \rv{X}_3^n,\rv{W}_3|\rv{W}_{c}) $ in the two following different ways:
\begin{align}
\Gamma(\rv{X}_1^n; \rv{X}_2^n; \rv{X}_3^n,\rv{W}_3|\rv{W}_{c})
&= H(\rv{X}_1^n|\rv{W}_c) + H(\rv{X}_2^n|\rv{W}_c) + H(\rv{X}_3^n,\rv{W}_3|\rv{W}_c) - H(\rv{X}_1^n,\rv{X}_2^n,\rv{X}_3^n,\rv{W}_3|\rv{W}_c) \\
&= H(\rv{X}_1^n|\rv{W}_c) + H(\rv{X}_2^n|\rv{W}_c) + H(\rv{W}_3|\rv{W}_c) - H(\rv{X}_1^n,\rv{X}_2^n,\rv{W}_3|\rv{W}_c) = \Gamma(\rv{X}_1^n; \rv{X}_2^n; \rv{W}_3|\rv{W}_{c})\\
&= \Gamma(\rv{X}_1^n; \rv{X}_2^n; \rv{X}_3^n|\rv{W}_{c}) + I(\rv{W}_3;\rv{X}_1^n,\rv{X}_2^n|\rv{W}_c,\rv{X}_3^n) \geq \Gamma(\rv{X}_1^n; \rv{X}_2^n; \rv{X}_3^n|\rv{W}_{c}).
\end{align}
In a very similar method, one can show that
\begin{align}
\Gamma(\rv{X}_1^n; \rv{X}_2^n; \rv{X}_3^n|\rv{W}_{c}) \leq 
\Gamma(\rv{W}_1; \rv{W}_2; \rv{W}_3|\rv{W}_{c}).
\end{align}
We proceed to derive an upper bound on $ \Gamma(\rv{W}_1; \rv{W}_2; \rv{W}_3|\rv{W}_{c}) $. Consider the subsequent list of inequalities:
\begin{align}
\Gamma(\rv{W}_1; \rv{W}_2; \rv{W}_3|\rv{W}_{c})
&= H(\rv{W}_1|\rv{W}_{c}) + H(\rv{W}_2|\rv{W}_{c})   - H(\rv{W}_1,\rv{W}_2|\rv{W}_3,\rv{W}_{c}) \\
&= H(\rv{W}_1|\rv{W}_{c}) + H(\rv{W}_2|\rv{W}_{c})   - H(\rv{W}_1,\rv{W}_2|\rv{W}_3,\rv{W}_{12},\rv{W}_{21},\rv{W}_{31},\rv{W}_{32}) \\
&\leq \Gamma(\rv{W}_1; \rv{W}_2; \rv{W}_3|\rv{W}_{12},\rv{W}_{21},\rv{W}_{31},\rv{W}_{32}) \\
&= H(\rv{W}_1|\rv{W}_3,\rv{W}_{12},\rv{W}_{21},\rv{W}_{31},\rv{W}_{32}) + H(\rv{W}_3|\rv{W}_3,\rv{W}_{12},\rv{W}_{21},\rv{W}_{31},\rv{W}_{32})   \\
&- H(\rv{W}_1,\rv{W}_3|\rv{W}_2,\rv{W}_{12},\rv{W}_{21},\rv{W}_{31},\rv{W}_{32}) \\
&= H(\rv{W}_1|\rv{W}_{12},\rv{W}_{21},\rv{W}_{31},\rv{W}_{32}) + H(\rv{W}_3|\rv{W}_{12},\rv{W}_{21},\rv{W}_{31},\rv{W}_{32})   \\
&- H(\rv{W}_1,\rv{W}_3|\rv{W}_2,\rv{W}_{21},\rv{W}_{31}) \\
&\leq \Gamma(\rv{W}_1; \rv{W}_2; \rv{W}_3|\rv{W}_{21},\rv{W}_{31}) \\ 
&= H(\rv{W}_2|\rv{W}_{21},\rv{W}_{31}) + H(\rv{W}_3|\rv{W}_{21},\rv{W}_{31}) - H(\rv{W}_2,\rv{W}_3|\rv{W}_1,\rv{W}_{21},\rv{W}_{31})   \\
&\leq \Gamma(\rv{W}_1; \rv{W}_2; \rv{W}_3) .
\end{align}
This completes the proof of the lemma.

	\subsection{Expected Size of Independently Generated Codebooks} \label{section:expected_size_of_codebooks}
	\begin{lemma} \label{lemma:expected_size_of_codebooks}
	Let $ (\rv{U},\rv{V}) \sim \mathsf{P}_{\rv{U}\rv{V}} $. Further, let $ \rv{V}^n $ be generated according to $ \prod_{i=1}^n \mathsf{P}_{\rv{V}}(v_i) $. Consider a codebook $ \set{C} = \{ \rv{U}^n(1),\dots \rv{U}^n(2^{nR})\} $. The codewords of $ \set{C} $ are generated independently each according to $ \prod_{i=1}^n \mathsf{P}_{\rv{U}}(u_i) $. Define the set
	\begin{equation}
		\set{D} = \{u^n: \in \set{C} : (u^n,\rv{V}^n) \in \set{T}_{\epsilon}^{(n)} (\rv{U},\rv{V})\}.
	\end{equation}
	Then, there exists $ \delta(\epsilon) > 0 $ that tends to zero as $ \epsilon \rightarrow 0 $ such that
	\begin{equation}
		\Exp{|\set{D}|} \leq 2^{n(R - I(\rv{U};\rv{V}) + \delta(\epsilon))}.
	\end{equation}
\end{lemma}
\begin{proof}
	Using definition of the set $ \set{D} $, we have
	\begin{align}
		|\set{D}| 
		&= \sum_{u^n  \in \set{C} } \indicator\{(u^n, \rv{V}^n) \in \set{T}_{\epsilon}^{(n)} (\rv{U},\rv{V})\} \\
		&= \sum_{m =1 }^{2^{nR}} \indicator\{(u^n(m),\rv{V}^n) \in \set{T}_{\epsilon}^{(n)} (\rv{U},\rv{V})\}.
	\end{align}
	Taking the expectation with respect to the codebook $ \set{C} $, and utilizing symmetry in codebook realization, we obtain
	\begin{align}
		\Exp{|\set{D}|}
		&= \sum_{m=1}^{2^{nR}}  \Prob{(\rv{U}^n(m), \rv{V}^n ) \in \set{T}_{\epsilon}^{(n)} (\rv{U},\rv{V})} \\
		&= 2^{nR} \cdot \Prob{(\rv{U}^n(1),\rv{V}^n) \in \set{T}_{\epsilon}^{(n)} (\rv{U},\rv{V})} \\
		&= 2^{nR}\sum_{(u^n,v^n) \in \set{T}_{\epsilon}^{(n)} (\rv{U},\rv{V})} \mathsf{P}_{\rv{U}^n}(u^n)\mathsf{P}_{\rv{V}^n}(v^n) \\
		&\leq 2^{nR}\sum_{(u^n,v^n) \in \set{T}_{\epsilon}^{(n)} (\rv{U},\rv{V})} 2^{-n(H(\rv{U}) +H( \rv{V}) ) - \delta(\epsilon))}\\
		&\leq 2^{nR} 2^{n(H(\rv{U,\rv{V}})+\delta(\epsilon))} 2^{-n(H(\rv{U}) +H( \rv{V})   - \delta(\epsilon))}.
	\end{align}
	This settles the proof of the lemma.
\end{proof}

	\subsection{Multivariate Covering Lemma}\label{section:covering_lemma}
	\begin{lemma}[Multivariate Covering Lemma]\label{lemma:covering_lemma}
	Denote $ \Omega $ to be an index set. Let $ \rv{U}_{\Omega}  \sim P_{\rv{U}_{\Omega}} $ and $ \epsilon > \epsilon' > 0 $. 
	For $ \omega \in \Omega $, randomly and independently generate sequences $ \rv{U}_{\Omega}^n (k_{\Omega}) $, $ k_{\omega} \in \setrn{R_{\omega}}{n} $, each according to $ \prod_{i=1}^n \mathsf{P}_{\rv{U}_{\omega}} (u_{\omega i}) $. Randomly and independently assign an index $ m(k_{\Omega}) $ to each index tuple $ k_{\Omega} \in \prod_{\omega \in \Omega }\setrn{R_{\omega}}{n}   $ according to a uniform \acrshort{pmf} over $ \setrn{R}{n} $. Denote each subset of index tuples with the same index $ m $ as a bin $ \mathcal{B}(m) $. Define for each tuple $ k_{\Omega} $,
	\begin{equation}
		\tilde{\mathcal{E}}(k_{\Omega}) \triangleq \{\rv{U}_{\Omega} ^n (k_{\Omega}) \notin \typsetn{\epsilon}{\rv{U}_{\Omega}} \},
	\end{equation}
	and for each $ m \in \setrn{R}{n} $ the event
	\begin{equation}
		\mathcal{E}(m) \triangleq \bigcap_{k_{\Omega} \in \set{B}(m)} \tilde{\mathcal{E}}(k_{\Omega}).
	\end{equation}
	Then, for each $ m $, there exists $ \delta(\epsilon) $ that tends to zero as $ \epsilon \rightarrow 0 $ such that $ \lim_{n\rightarrow \infty} \Prob{\mathcal{E}(m)} = 0 $, if
	\begin{equation}
		\sum_{\omega \in \set{S}} R_{\omega} > \indicator \{\set{S} = \Omega\} \cdot  R + \Gamma(\rv{U}_{\set{S}}) ,
	\end{equation}
	for all $ \set{S} \subseteq \Omega  $, such that $ |\set{S}| \geq 2 $.
	
\end{lemma}

\begin{proof}
	By symmetry, it suffices to investigate the case $ m=1 $.
	Let
	\begin{equation}
		\mathcal{A} 
		\triangleq 
		\{ k_{\Omega} \in \mathcal{B}(1)  : \rv{U}_{\Omega}^n (k_{\Omega}) \in \typsetn{\epsilon}{\rv{U}_{\Omega}} \}.
	\end{equation}
	For convenience, denote
	\begin{equation}
		\phi(k_{\Omega}) \triangleq  \indicator \{  \rv{U}_{\Omega}^n (k_{\Omega}) \in \typsetn{\epsilon}{\rv{U}_{\Omega}} \}.
	\end{equation}
	Then, the set size $ |\mathcal{A}| $ conditioned on the random bin assignment $ \mathcal{B}(1) $ can be expressed as
	\begin{equation}
		|\mathcal{A}|_{| \mathcal{B}(1)} = \sum_{k_{\Omega} \in \mathcal{B}(1)  }  \phi(k_{\Omega}) .
	\end{equation}
	For $ a_{\Omega} \in \{1,2\}^{|\Omega|} $, let
	\begin{equation}
		p(a_{\Omega} ) \triangleq \Exp{\phi(1_{\Omega }) \phi(a_{\Omega} ) } ,
	\end{equation}
	and
	\begin{equation}
		Q(a_{\Omega}) \triangleq | \{(k_{\Omega},k_{\Omega}' )  : k_{\Omega} \in \mathcal{B}(1), k'_{\Omega} \in \mathcal{B}(1) , \mathcal{F}_{\Omega} ^{(a_{\Omega})} \},
	\end{equation}
	where $ \set{F}_{\omega}^{(1)} $ implies that $ k_{\omega} = k'_{\omega} $, and $ \set{F}_{\omega}^{(2)} = \set{F}_{\omega}^{(1)c} $ implies that $ k_{\omega} \neq k'_{\omega} $.
	Then, we have
	\begin{equation}
		\Exp{ |\mathcal{A}| \big | \mathcal{B}(1)} = \sum_{k_{\Omega} \in \mathcal{B}(1)  }  \Exp{\phi(k_{\Omega}) } = Q(1_{\Omega}) p(1_{\Omega}),
	\end{equation}
	and
	\begin{align}
		\Exp{ |\mathcal{A}|^2 \big | \mathcal{B}(1)} 
		=& \sum_{k_{\Omega} \in \mathcal{B}(1)  } \sum_{k'_{\Omega} \in \mathcal{B}(1)  }  \Exp{\phi(k_{\Omega})  \phi(k'_{\Omega}) } \\
		&= \sum_{ a_{\Omega} \in \{1,2\}^{|\Omega}} Q(a_{\Omega}) p(a_{\Omega} ) .
	\end{align}
	Note that by the joint typicality lemma, it follows that
	\begin{align}
		p(1_{\Omega}) 
		&=   \Exp{\phi(1_{\Omega }) \phi(1_{\Omega} ) } \\
		&= \Prob{\rv{U}_{\Omega}^n (1_{\Omega}) \in \mathcal{T}_{\epsilon}^{(n)} (\mathsf{P}_{\rv{U}_{\Omega}})} \\
		&\geq 2^{-n(\Gamma(\rv{U}_{\Omega}) + \delta(\epsilon))},
	\end{align}
	and
	\begin{align}
		p(a_{\Omega} ) 
		&= \Exp{\phi(1_{\Omega}) \phi(a_{\Omega} ) } \\
		&= \Prob{\rv{U}_{\Omega}^n (1_{\Omega}) \in \mathcal{T}_{\epsilon}^{(n)} (\mathsf{P}_{\rv{U}_{\Omega}}), \rv{U}_{\Omega}^n (a_{\Omega}) \in \mathcal{T}_{\epsilon}^{(n)} (\mathsf{P}_{\rv{U}_{\Omega}})} \\
		&= \Prob{\rv{U}_{\Omega}^n (1_{\Omega}) \in \mathcal{T}_{\epsilon}^{(n)} (\mathsf{P}_{\rv{U}_{\Omega}}), (\rv{U}_{\mathcal{S}}^n (1_{\mathcal{S}}) , \rv{U}_{\mathcal{S}^c}^n (2_{\mathcal{S}^c})) \in \mathcal{T}_{\epsilon}^{(n)} (\mathsf{P}_{\rv{U}_{\Omega}})} \\
		&= \sum_{u^n_{\Omega} \in \mathcal{T}_{\epsilon}^{(n)} (\rv{U}_{\Omega})} 
		\mathsf{P}_{\rv{U}^n_{\Omega}} (u^n_{\Omega})
		\sum_{u^n_{\mathcal{S}^c} \in \mathcal{T}_{\epsilon}^{(n)} (\rv{U}_{\mathcal{S}^c}|u^n_{\mathcal{S}})}  P_{\rv{U}^n_{\mathcal{S}^c}} (u^n_{\mathcal{S}^c})\\
		&\leq 2^{-n(\Gamma(\rv{U}_{\Omega})  + \sum_{\omega \in \mathcal{S}^c}  H(\rv{U}_{\omega}) - H(\rv{U}_{\mathcal{S}^c} | \rv{U}_{\mathcal{S}}) -\delta(\epsilon) )}.
	\end{align}
	Also, for all $ a_{\Omega} \in \{1,2\}^{|\Omega|} $, we have
	\begin{align}
		\Exp{Q(a_{\Omega}) }
		&= 2^{n(\sum_{\omega \in \Omega} a_{\omega} R_{\omega}- (1+\indicator \{\bigcup_{\omega \in \Omega} a_{\omega} = 2\}) R)}.
	\end{align}
	Finally
	\begin{align}
		\Prob{\mathcal{E}(1)} 
		&=\frac{\Exp{\Exp{|\set{A}|^2 |\set{B}(1)}} - \left(\Exp{\Exp{|\set{A}|| \set{B}(1)}} \right)^2}{\left(\Exp{\Exp{|\set{A}|| \set{B}(1)}} \right)^2} \\
		&=\frac{\sum_{ a_{\Omega} \in \{1,2\}^{|\Omega}|} \Exp{ Q(a_{\Omega})} p(a_{\Omega} )  - \left(\Exp{Q(1_{\Omega})} p(1_{\Omega}) \right)^2}{\left(\Exp{Q(1_{\Omega})} p(1_{\Omega}) \right)^2} \\
		&=\frac{\sum_{ a_{\Omega} \in \{1,2\}^{|\Omega}|/ 2^{|\Omega|}} \Exp{ Q(a_{\Omega})} p(a_{\Omega} ) }{\left(\Exp{Q(1_{\Omega})} p(1_{\Omega}) \right)^2} \\
		&\leq \sum_{ a_{\Omega} \in \{1,2\}^{|\Omega}|/ 2^{|\Omega|}} \frac{2^{n(\sum_{\omega \in \Omega} a_{\omega} R_{\omega}- (1+\indicator \{\bigcup_{\omega \in \Omega} a_{\omega} = 2\}) R)} 2^{-n(\Gamma(\rv{U}_{\Omega})  + \sum_{\omega \in \mathcal{S}^c}  H(\rv{U}_{\omega}) - H(\rv{U}_{\mathcal{S}^c} | \rv{U}_{\mathcal{S}}) -\delta(\epsilon) )}}{2^{n(\sum_{\omega \in \Omega} 2 R_{\omega}- 2 R)} 2^{-n(2\Gamma(\rv{U}_{\Omega}) + 2\delta(\epsilon))}} \\
		&= \sum_{ a_{\Omega} \in \{1,2\}^{|\Omega}|/ 2^{|\Omega|}} 2^{-n( -\sum_{\omega \in \Omega} a_{\omega} R_{\omega}+ (1+\indicator \{\bigcup_{\omega \in \Omega} a_{\omega} = 2\}) R + \Gamma(\rv{U}_{\Omega})  + \sum_{\omega \in \mathcal{S}^c}  H(\rv{U}_{\omega}) - H(\rv{U}_{\mathcal{S}^c} | \rv{U}_{\mathcal{S}}) -\delta(\epsilon) + \sum_{\omega \in \Omega} 2 R_{\omega} -  2 R -2 \Gamma(\rv{U}_{\Omega}) - 2\delta(\epsilon))} \\
		&= \sum_{ a_{\Omega} \in \{1,2\}^{|\Omega}|/ 2^{|\Omega|}} 2^{-n( \sum_{\omega \in \Omega} \indicator \{ a_{\omega} = 1\}  R_{\omega}-  \indicator \{\bigcap_{\omega \in \Omega} a_{\omega} = 1\}) R - \Gamma(\rv{U}_{\Omega})  + \sum_{\omega \in \mathcal{S}^c}  H(\rv{U}_{\omega}) - H(\rv{U}_{\mathcal{S}^c} | \rv{U}_{\mathcal{S}})    - 3\delta(\epsilon))} \\
		&= \sum_{ a_{\Omega} \in \{1,2\}^{|\Omega}|/ 2^{|\Omega|}} 2^{-n( \sum_{\omega \in \Omega} \indicator \{ a_{\omega} = 1\}  R_{\omega}-  \indicator \{\bigcap_{\omega \in \Omega} a_{\omega} = 1\}) R - \sum_{\omega \in \Omega} H(\rv{U}_{\omega}) + H(\rv{U}_{\Omega})  + \sum_{\omega \in \mathcal{S}^c}  H(\rv{U}_{\omega}) - H(\rv{U}_{\mathcal{S}^c} | \rv{U}_{\mathcal{S}})    - 3\delta(\epsilon))} \\
		&= \sum_{ a_{\Omega} \in \{1,2\}^{|\Omega}|/ 2^{|\Omega|}} 2^{-n( \sum_{\omega \in \Omega} \indicator \{ a_{\omega} = 1\}  R_{\omega}-  \indicator \{\bigcap_{\omega \in \Omega} a_{\omega} = 1\}) R - \sum_{\omega \in \set{S}} H(\rv{U}_{\omega}) + H(\rv{U}_{\set{S}})     - 3\delta(\epsilon))} \\
		&= \sum_{ a_{\Omega} \in \{1,2\}^{|\Omega}|/ 2^{|\Omega|}} 2^{-n( \sum_{\omega \in \Omega} \indicator \{ a_{\omega} = 1\}  R_{\omega}-  \indicator \{\bigcap_{\omega \in \Omega} a_{\omega} = 1\}) R -  \Gamma(\rv{U}_{\set{S}})     - 3\delta(\epsilon))}.
	\end{align}
	This completes the proof of the lemma.
\end{proof}

	\subsection{Multivariate Packing Lemma}\label{section:packing_lemma}
	\begin{lemma} \label{lemma:packing_lemma}
	Let $ (\rv{U}_{\set{S}}, \rv{U}_{\set{S}^c},\rv{Y}) \sim \mathsf{P}_{\rv{U}_{\set{S}}\rv{U}_{\set{S}^c}\rv{Y}} $. Let $ (\rv{U}_{\set{S}^c}^n, \rv{Y}^n) \sim \mathsf{P}_{\rv{U}_{\set{S}^c}^n \rv{Y}^n} $ be a tuple of arbitrarily distributed random sequences, not necessarily distributed according to $ \prod_{i=1}^n \mathsf{P}_{\rv{U}\rv{Y}} (u_{\Omega,i},y_i) $. Let $ \rv{U}_{\set{S}}^n(k_{\set{S}}) $, $ k_{\set{S}} \in \mathcal{A} $, where $ |\mathcal{A}| \leq 2^{n (\sum_{\omega \in \set{S} } R_{\omega})} $, be random sequences, each distributed according to $ \prod_{i=1}^n \mathsf{P}_{\rv{U}_{\omega}} (u_{\omega,i}) $, for each $ \omega \in \set{S} $. Furthermore, assume that $ \rv{U}_{\set{S}}^n(k_{\set{S}}) $, $ k_{\set{S}} \in \mathcal{A} $, is pairwise conditionally independent of $ ( \rv{U}_{\set{S}^c}^n, \rv{Y}^n) $ , but is arbitrarily dependent on other $ \rv{U}_{\set{S}}^n(k_{\set{S}}) $ sequences. Then, there exists $ \delta(\epsilon) $ that tends to zero as $ \epsilon \rightarrow 0 $ such that
	\begin{equation}
		\lim_{n\rightarrow \infty} \Prob{(\rv{U}_{\set{S}}^n (k_{\set{S}}), \rv{U}_{\set{S}^c}^n, \rv{Y}^n)\in \typsetn{\epsilon}{ \rv{U}_{\set{S}}, \rv{U}_{\set{S}^c}, \rv{Y}} \text{ for some } k_{\set{S}} \in \mathcal{A}} =0
	\end{equation}
	if 
	\begin{equation}
		\sum_{\omega \in \set{S}} R_{\omega} < I(\rv{U}_{\set{S}}; \rv{U}_{\set{S}^c},\rv{Y}) + \Gamma (\rv{U}_{\set{S}})- \delta(\epsilon).
	\end{equation}
\end{lemma}
\begin{proof}
	Following standard techniques in probability, we have
	\begin{align}
		&\Prob{(\rv{U}^n_{\set{S}}(k_{\set{S}}), \rv{U}^n_{\set{S}^c} ,\rv{Y}^n) \in \typsetn{\epsilon}{ \rv{U}_{\set{S}}, \rv{U}_{\set{S}^c}, \rv{Y}}  } \\
		&= \sum_{(u^n_{\set{S}^c} ,y^n) \in \mathcal{T}_{\epsilon}^{(n) } (\rv{U}^n_{\set{S}},\rv{Y})} P_{\rv{U}^n_{\set{S}^c} \rv{Y}^n}(u^n_{\set{S}_c},y^n) \Prob{(\rv{U}^n_{\set{S}}(k_{\set{S}}), \rv{U}^n_{\set{S}^c} ,\rv{Y}^n) \in \mathcal{T}_{\epsilon}^{(n)} (\rv{U}_{\set{S}}, \rv{U}_{\set{S}^c}, \rv{Y}) \big| \rv{U}^n_{\set{S}^c} = u^n_{\set{S}^c} , \rv{Y}^n = y^n} \\
		&= \sum_{(u^n_{\set{S}^c} ,y^n) \in \mathcal{T}_{\epsilon}^{(n) } (\rv{U}^n_{\set{S}},\rv{Y})} P_{\rv{U}^n_{\set{S}^c} \rv{Y}^n}(u^n_{\set{S}_c},y^n) \Prob{(\rv{U}^n_{\set{S}}(k_{\set{S}}), u^n_{\set{S}^c} ,y^n) \in \mathcal{T}_{\epsilon}^{(n)} (\rv{U}_{\set{S}}, \rv{U}_{\set{S}^c}, \rv{Y})  }.
	\end{align}
	The probability of the event in the sum can be bounded from above as follows:
	\begin{align}
		\Prob{(\rv{U}^n_{\set{S}}(k_{\set{S}}), u^n_{\set{S}^c} ,y^n) \in \mathcal{T}_{\epsilon}^{(n)} (\rv{U}_{\set{S}}, \rv{U}_{\set{S}^c}, \rv{Y})  }
		&= \sum_{u^n_{\set{S}} \in \mathcal{T}_{\epsilon}^{(n)} (\rv{U}_{\set{S}}|u^n_{\set{S}^c},y^n)} \mathsf{P}_{\rv{U}^n_{\set{S}}} (u^n_{\set{S}}) \\
		&\leq |\mathcal{T}_{\epsilon}^{(n)} (\rv{U}_{\set{S}}|u^n_{\set{S}^c},y^n)| 2^{-n(\sum_{i \in \set{S}} H(\rv{U}_i) - \delta(\epsilon))},
	\end{align}
	where the size of conditional typical set can be upper bounded as
	\begin{align}
		1
		&= \sum_{u^n_{\set{S}} \in \mathcal{U}^n_{\set{S}}} \mathsf{P}_{\rv{U}^n_{\set{S}}|\rv{U}^n_{\set{S}^c} \rv{Y}^n } (u^n_{\set{S}}|u^n_{\set{S}^c},y^n) \\
		&\geq \sum_{u^n_{\set{S}} \in \mathcal{T}_{\epsilon}^{(n)} (\rv{U}_{\set{S}}|u^n_{\set{S}^c},y^n)} \mathsf{P}_{\rv{U}^n_{\set{S}}|\rv{U}^n_{\set{S}^c} \rv{Y}^n } (u^n_{\set{S}}|u^n_{\set{S}^c},y^n)\\
		&\geq |\mathcal{T}_{\epsilon}^{(n)} (\rv{U}_{\set{S}}|u^n_{\set{S}^c},y^n)| 2^{-n(H(\rv{U}_{\set{S}}|\rv{U}_{\set{S}^c} ,\rv{Y}) +\delta(\epsilon))}.
	\end{align}
	Therefore,
	\begin{equation}
		\Prob{(\rv{U}^n_{\set{S}}(k_{\set{S}}), u^n_{\set{S}^c},y^n) \in \mathcal{T}_{\epsilon}^{(n)} (\rv{U}_{\set{S}}, \rv{U}_{\set{S}^c}, \rv{Y})  } \leq 
		2^{-n(\sum_{i \in \set{S}} H(\rv{U}_i) - H(\rv{U}_{\set{S}}|\rv{U}_{\set{S}^c} ,\rv{Y})  - \delta(\epsilon))},
	\end{equation}
	and
	\begin{equation}
		\Prob{(\rv{U}^n_{\set{S}}(k_{\set{S}}), \rv{U}^n_{\set{S}^c} ,\rv{Y}^n) \in \mathcal{T}_{\epsilon}^{(n)} (\mathsf{P}_{\rv{U}_{\set{S}} \rv{U}_{\set{S}^c} \rv{Y}})} \leq 2^{-n(\sum_{i \in \set{S}} H(\rv{U}_i) - H(\rv{U}_{\set{S}}|\rv{U}_{\set{S}^c} ,\rv{Y})  - \delta(\epsilon))}.
	\end{equation}
	The probability of the event of interest can be further bounded as follows
	\begin{align}
		&\Prob{(\rv{U}^n_{\set{S}}(k_{\set{S}}), \rv{U}^n_{\set{S}^c},\rv{Y}^n) \in \mathcal{T}_{\epsilon}^{(n)} (\mathsf{P}_{\rv{U}_{\set{S}} \rv{U}_{\set{S}^c} \rv{Y}}) \text{ for some } k_{\set{S}} \in \setrn{R_{\set{S}}}{n}/ K_{\set{S}}} \\
		&=\Prob{ \bigcap_{k_{\set{S}} \in \setrn{R_{\set{S}}}{n}/ K_{\set{S}}} (\rv{U}^n_{\set{S}}(k_{\set{S}}), \rv{U}^n_{\set{S}^c},\rv{Y}^n) \in \mathcal{T}_{\epsilon}^{(n)} (\mathsf{P}_{\rv{U}_{\set{S}} \rv{U}_{\set{S}^c} \rv{Y}}) } \\
		&=\prod_{k_{\set{S}} \in \setrn{R_{\set{S}}}{n}/ K_{\set{S}}} \Prob{  (\rv{U}^n_{\set{S}}(k_{\set{S}}), \rv{U}^n_{\set{S}^c} ,\rv{Y}^n) \in \mathcal{T}_{\epsilon}^{(n)} (\mathsf{P}_{\rv{U}_{\set{S}} \rv{U}_{\set{S}^c} \rv{Y}}) } \\
		&\leq 2^{-n(\sum_{i \in \set{S}} H(\rv{U}_i) - H(\rv{U}_{\set{S}}|\rv{U}_{\set{S}^c} ,\rv{Y}) - \sum_{\omega \in \set{S}} R_\omega  - \delta(\epsilon))}.
	\end{align}
	This concludes the proof of the packing lemma.
	
\end{proof}

	\subsection{Properties of Total Correlation}\label{section:total_correlation_properties}
	In this section we will list and prove various properties concerning the total correlation among random variables, i.e.:
\begin{equation}
    \Gamma(\rv{X}_{\set{S}}) \triangleq \sum_{\omega \in \set{S}} H(\rv{X}_{\omega}) - H(\rv{X}_{\set{S}}).
\end{equation}
\begin{proposition}[Properties of Total Correlation] \label{proposition:total_correlation_properties}
	Consider a set of random variables $ \rv{X}_{\set{S}} $, where $ \set{S} $ is some index set. Let $ K \triangleq |\set{S}| $ be the size of the set $ \set{S} $. We enumerate the elements of $ \set{S} $ using an index from $ [K] = \{1,2,\dots,K\} $. For simplicity we consider an equivalent set of random variables $ \rv{X}_{[|K|]} $. The total correlation of  $ \rv{X}_{[|\set{S}|]} $ resembles the following list of properties:
	\begin{enumerate}
		\item 
		\begin{equation}
			\Gamma(\rv{X}_{[K]}) = \sum_{k=1}^K I(\rv{X}_k;\rv{X}_{[k-1]}).
		\end{equation}
		\item 
		\begin{equation}
			\Gamma (\rv{X}_1^k) =  \Gamma(\rv{X}_1^{k-1}) + I(\rv{X}_k; \rv{X}_1^{k-1})
		\end{equation}
	\end{enumerate}
\end{proposition}
\begin{proof}
	\begin{enumerate}
		\item By definition
		\begin{align*}
			\Gamma(\rv{X}_{[K]})
			&= \sum_{k=1}^K H(\rv{X}_k) - H(\rv{X}_{[K]}) \\
			&= \sum_{k=1}^{K-1} H(\rv{X}_k) - H(\rv{X}_{[K-1]}) + H(\rv{X}_{K}) - H(\rv{X}_K|\rv{X}_{[K-1]}) \\
			&= \Gamma(\rv{X}_{[K-1]}) + I(\rv{X}_K;\rv{X}_{[K-1]}) \\
			&= \sum_{k=1}^K I(\rv{X}_k;\rv{X}_{[k-1]}).
		\end{align*}
		\item 
		\begin{align}
			\Gamma (\rv{X}_1^k)
			&= \sum_{i=1}^{k} H(\rv{X}_i) - H (\rv{X}_1^k) \\
			&= \sum_{i=1}^{k-1} H(\rv{X}_i) - H (\rv{X}_1^{k-1}) + H(\rv{X}_k) - H(\rv{X}_k|\rv{X}_{1}^{k-1}) \\
			&= \Gamma(\rv{X}_1^{k-1}) + I(\rv{X}_k; \rv{X}_1^{k-1})
		\end{align}
	\end{enumerate}
\end{proof}

	\subsection{Bounds on Differential Entropy}\label{section:differential_entropy_bounds}
	\begin{proposition} \label{proposition:conditional_differential_entropy_upper_bound}
	Let $ \rv{X} $ be a continuous random variable and $ \bv{Y} $ some random vector  with a joint probability distribution function $ F_{\rv{X}\bv{Y}} $. Denote:
	\begin{align}
		\mu_x &\triangleq \Exp{\rv{X}}, \\
		\sigma_x^2 &\triangleq \var{\rv{X}} = \Exp{(\rv{X}-\Exp{\rv{X}})^2}, \\
		\boldsymbol{\mu}_y & \triangleq \Exp{\bv{Y}}, \\
		\Sigma_y &\triangleq \Exp{\bv{Y}\bv{Y}^T} - \Exp{\bv{Y}} \Exp{\bv{Y}}^T ,\\
		\Sigma_{\rv{X} \bv{Y} } &\triangleq \cov{\rv{X},\bv{Y}}= \Exp{\rv{X} \bv{Y}^T} -\Exp{\rv{X}} \Exp{\bv{Y}}^T.
	\end{align}
	  The conditional differential entropy of $ \rv{X} $ given $ \rv{Y} $ is bounded from above as follows:
	\begin{equation}
		h(\rv{X}|\bv{Y}) \leq \frac{1}{2} \log \left(2\pi e \left(\sigma_x^2 - \Sigma_{\rv{X} \bv{Y} } \Sigma_{\bv{Y}}^{-1} \Sigma_{\rv{X} \bv{Y} }^T\right) \right).
	\end{equation}
\end{proposition}
\begin{proof}
	For any function of $ \bv{Y} $, $ f(\bv{Y}) $,
	\begin{align}
		h(\rv{X}|\bv{Y})
		&= h(\rv{X} - f(\bv{Y})|\bv{Y}) \\
		&\eqann{\leq}{a} h(\rv{X} - f(\bv{Y})) \\
		&\eqann{\leq}{b} \frac{1}{2} \log \left(2\pi e \Exp{(\rv{X}-f(\bv{Y}))^2}\right),
	\end{align}
	where \eqannref{a} follows since conditioning reduces differential entropy \cite[Sec. 8.6]{Cover2006}, and \eqannref{b} holds since Gaussian random variables maximize differential entropy with bounded variance. Now we choose $ f(\bv{Y}) $ to be the linear MMSE estimator of $ \rv{X} $, i.e.,
	\begin{equation}
		f(\bv{Y})  = \hat{\rv{X}}_{lin}(\bv{Y}) =\mu_x + \Sigma_{\rv{X} \bv{Y} } \Sigma_{\bv{Y}}^{-1} (\bv{Y} - \boldsymbol{\mu}_y).
	\end{equation}
	By the orthogonality principle
	\begin{align}
		\Exp{(\rv{X}-\hat{\rv{X}}_{lin}(\bv{Y}) )^2}
		&= \Exp{(\rv{X}-\hat{\rv{X}}_{lin}(\bv{Y}) ) \rv{X}} \\
		&= \Exp{\rv{X}^2} - \Exp{\mu_x \rv{X} + \Sigma_{\rv{X} \bv{Y} } \Sigma_{\bv{Y}}^{-1} (\bv{Y} \rv{X} - \boldsymbol{\mu}_y \rv{X})} \\
		&= \sigma_x^2 - \Sigma_{\rv{X} \bv{Y} } \Sigma_{\bv{Y}}^{-1} \Sigma_{\rv{X} \bv{Y} }^T.
	\end{align}
\end{proof}

	\bibliographystyle{IEEEtran}
	\bibliography{./cran}
\end{document}